\documentclass[grad,pdftex]{poli}
\usepackage[utf8]{inputenc}
\usepackage{float}
\usepackage{multirow}
\usepackage{wrapfig}
\usepackage{textgreek}
\usepackage{titlesec}
\usepackage{comment}
\usepackage{blindtext}
\usepackage{ragged2e}

\usepackage[T1]{fontenc}

\usepackage{lmodern}
\usepackage{ae,aecompl}								
\usepackage[upright]{fourier}

\usepackage{enumerate}
\usepackage{enumitem}
\usepackage[section]{placeins}

\usepackage{epigraph}
\usepackage[font={small}]{caption}
\usepackage[nohints]{minitoc}

\usepackage{amsmath}
\usepackage{amssymb}
\usepackage{amsfonts}
\usepackage{nicefrac}

\usepackage{multirow}
\usepackage{booktabs}
\usepackage{colortbl}
\usepackage{tabularx}
\usepackage{multirow}
\usepackage{threeparttable}
\usepackage{etoolbox}
\appto\TPTnoteSettings{\footnotesize}

\usepackage{graphicx}
\usepackage{subcaption}
\usepackage{pdfpages}
\usepackage{rotating}
\usepackage{pgfplots}
	\usepgfplotslibrary{groupplots}

\usepackage{tikz}
	\usetikzlibrary{backgrounds,automata}
	\pgfplotsset{width=7cm,compat=1.3}
	\tikzset{every picture/.style={execute at begin picture={
   		\shorthandoff{:;!?};}
	}}
	\pgfplotsset{every linear axis/.append style={
		/pgf/number format/.cd,
		use comma,
		1000 sep={\,},
	}}
\usepackage{eso-pic}
\usepackage{import}

\usepackage{xspace}
\usepackage{siunitx}
	\DeclareSIUnit{\MPa}{\mega\pascal}
	\DeclareSIUnit{\micron}{\micro\meter}
	\DeclareSIUnit{\tr}{tr}
	\DeclareSIPostPower\totheM{m}
\usepackage[version=3]{mhchem}
\usepackage{textcomp}
\usepackage{array}
\usepackage{hyphenat}

\definecolor{coolblack}{rgb}{0.0, 0.18, 0.4}

\usepackage[pdftex,pdfborder={0 0 0},
			colorlinks=true,
			linkcolor=coolblack,
			citecolor=red,
			pagebackref=true,
			]{hyperref}

\usepackage[top=2.5cm, bottom=2cm, left=3cm, right=2.5cm,
			headheight=15pt]{geometry}
			
\usepackage{fancyhdr}
\usepackage[acronym]{glossaries}
	\newglossary[nlg]{notation}{not}{ntn}{Notation}
	\makenoidxglossaries

\newacronym{CFD}{CFD}{Computational Fluid Dynamics}

\newacronym{DL}{DL}{Deep Learning}

\newacronym{GDL}{GDL}{Geometric Deep Learning}

\newacronym{aoa}{AoA}{Angle of Attack}

\newacronym{NS}{NS}{Navier-Stokes}

\newacronym{RANS}{RANS}{Reynolds Averaged Navier-Stokes}

\newacronym{PDE}{PDE}{Partial Differential Equation}

\newacronym{INR}{INR}{Implicit Neural Representation}

\newacronym{PINN}{PINN}{Physics Informed Neural Network}

\newacronym{AD}{AD}{Automatic Differentiation}

\newacronym{GNN}{GNN}{Graph Neural Network}

\newacronym{CNN}{CNN}{Convolution Neural Network}

\newacronym{HPC}{HPC}{High Performance Computing}

\newacronym{GNS}{GNS}{Graph Network-based Simulators}

\newacronym{LES}{LES}{Large Eddy Simulations}

\newacronym{DNS}{DNS}{Direct Numerical Simulation}

\newacronym{GAT}{GAT}{Graph Attention Network}

\newacronym{SIREN}{SIREN}{Sinusoidal Representation Networks}

\newacronym{MLP}{MLP}{Multi Layer Perceptron}

\newglossaryentry{p}{
  type=notation,
  name={\ensuremath{p}},
  description={pressure},
  sort={pressure}%
}

\newglossaryentry{x}{
  type=notation,
  name={\ensuremath{x}},
  description={Input data or design variables},
  sort={x}%
}

\newglossaryentry{hatx}{
  type=notation,
  name={\ensuremath{\hat{x}}},
  description={Scaled input data},
  sort={x}%
}

\newglossaryentry{y}{
  type=notation,
  name={\ensuremath{y}},
  description={Output data},
  sort={y}%
}

\newglossaryentry{bary}{
  type=notation,
  name={\ensuremath{\bar{y}}},
  description={Ground truth target label},
  sort={y}%
}

\newglossaryentry{haty}{
  type=notation,
  name={\ensuremath{\hat{y}}},
  description={Ground truth target label scaled},
  sort={y}%
}

\newglossaryentry{y+}{
  type=notation,
  name={\ensuremath{y+}},
  description={Non-dimensional wall distance for a wall-bounded flow},
  sort={yplus}%
}

\newglossaryentry{i}{
  type=notation,
  name={\ensuremath{i}},
  description={subscript used for a single sample from a dataset},
  sort={y}%
}

\newglossaryentry{U}{
  type=notation,
  name={$U$},
  description={Freestream velocity},
  sort={U}%
}

\newglossaryentry{$u$}{
  type=notation,
  name={$u$},
  description={Velocity in x axis},
  sort={u}%
}

\newglossaryentry{$v$}{
  type=notation,
  name={$v$},
  description={Velocity in y axis},
  sort={v}%
}

\newglossaryentry{rho}{
  type=notation,
  name={\ensuremath{\rho}},
  description={Fluid density},
  sort={rho}%
}

\newglossaryentry{nu}{
  type=notation,
  name={\ensuremath{\nu}},
  description={Fluid viscosity},
  sort={viscosity}%
}

\newglossaryentry{nut}{
  type=notation,
  name={\ensuremath{\nu_t}},
  description={Fluid turbulent viscosity},
  sort={viscosity}%
}

\newglossaryentry{alpha}{
  type=notation,
  name={\ensuremath{\alpha}},
  description={Scaling parameter used internally for gradient loss},
  sort={alpha}%
}

\newglossaryentry{gamma}{
  type=notation,
  name={\ensuremath{\gamma}},
  description={Scaling parameter between normal and gradient losses.},
  sort={gamma}%
}

\newglossaryentry{epsilon}{
  type=notation,
  name={\ensuremath{\epsilon}},
  description={Numerical step},
  sort={gamma}%
}

\newglossaryentry{n}{
  type=notation,
  name={\ensuremath{n}},
  description={Column index of a matrix},
  sort={n}%
}

\newglossaryentry{m}{
  type=notation,
  name={\ensuremath{m}},
  description={Row index of a matrix},
  sort={m}%
}

\newglossaryentry{em}{
  type=notation,
  name={\ensuremath{e_m}},
  description={Unit vector},
  sort={e}%
}

\newglossaryentry{M}{
  type=notation,
  name={\ensuremath{M}},
  description={Maximum ordinate of the camber line},
  sort={M}%
}

\newglossaryentry{P}{
  type=notation,
  name={\ensuremath{P}},
  description={Distance of the furthest point from the camber line, to the leading edge},
  sort={P}%
}

\newglossaryentry{XX}{
  type=notation,
  name={\ensuremath{XX}},
  description={Maximum thickness},
  sort={XX}%
}

\newglossaryentry{mu}{
  type=notation,
  name={\ensuremath{\bar{\mu}}},
  description={Mean from ground truth data},
  sort={mu}%
}

\newglossaryentry{sigma}{
  type=notation,
  name={\ensuremath{\bar{\sigma}}},
  description={Standard deviation},
  sort={sigma}%
}

\newglossaryentry{LG}{
  type=notation,
  name={\ensuremath{L_G}},
  description={Gradient loss function},
  sort={LG}%
}

\newglossaryentry{LN}{
  type=notation,
  name={\ensuremath{L_N}},
  description={Normal loss function},
  sort={LN}%
}

\input{Configs/FirstPage}

  

\begin{document}
  \title{Using Gradient to Boost Generalization Performance of Deep Learning Models for Fluid Dynamics}
  \foreigntitle{Utilisation de Gradient pour Améliorer la Généralisation de Modèles de Deep Learning pour la Dynamique de Fluides}
 \author{Eduardo}{Vital Brasil Lorenzo Fernandez}
 \date{7}{10}{2022}
  \department{CEMEF}
  \advisor{Thibaut Munzer}
  \director{Youssef Mesri}
  \engschool{MINES Paristech}

  \hypersetup{
    pdfauthor={Vital Brasil Eduardo},
    pdfsubject={Thesis},
    pdftitle={Using Gradient to Boost Generalization Performance of Deep Learning Models for Fluid Dynamics},
    pdfkeywords={Deep Learning, CFD, Adjoint Method}
}

 \definecolor{gray75}{gray}{0.75}
  \newcommand{\hsp}{\hspace{20pt}}
  \titleformat{\chapter}[hang]{\Huge\bfseries}{\thechapter\hsp\textcolor{gray75}{|}\hsp}{0pt}{\Huge\bfseries}

  \pagenumbering{roman}
  \firstpage
  \cleardoublepage 
  \begin{abstract}

Nowadays, \gls{CFD} is a fundamental tool for industrial design. However, the computational cost of doing such simulations is expensive and can be detrimental for real-world use cases where many simulations are necessary, such as the task of shape optimization. Recently, \gls{DL} has achieved a significant leap in a wide spectrum of applications and became a good candidate for physical systems, opening perspectives to CFD. To circumvent the computational bottleneck of CFD, DL models have been used to learn on Euclidean data, and more recently, on non-Euclidean data such as unstuctured grids and manifolds, allowing much faster and more efficient (memory, hardware) surrogate models. Nevertheless, DL presents the intrinsic limitation of extrapolating (generalizing) out of training data distribution (design space). In this study, we present a novel work to increase the generalization capabilities of Deep Learning. To do so, we incorporate the physical gradients (derivatives of the outputs \textit{w.r.t.} the inputs) to the DL models. Our strategy has shown good results towards a better generalization of DL networks and our methodological/ theoretical study is corroborated with empirical validation, including an ablation study.

\end{abstract}

  \begin{foreignabstract}

Actuellement, la CFD est un outil fondamental pour la conception industrielle. Cependant, le coût de calcul de telles simulations est élevé et peut être préjudiciable dans les cas d’utilisation réelles, où de nombreuses simulations sont nécessaires, comme la tâche de l’optimisation de forme. Récemment, le DL a attendu un important avancement dans un large spectre d’applications et est devenu un bon candidat pour des systèmes physiques, ouvrant des perspectives pour la CFD. Afin de débloquer cette limitation numérique, des modèles de DL ont été utilisés pour apprendre sur les données euclidiennes et, plus récemment, sur les données non euclidiennes telles que les graphs et les \textit{manifolds}, permettant des modèles substituts beaucoup plus rapides et efficaces (mémoire, hardware). Néanmoins, le DL présente la limitation intrinsèque de l’extrapolation (généralisation) hors la distribution des données d'entrainement (espace de conception). Dans cette étude, nous présentons un nouveau travail visant à accroître les capacités de généralisation du Deep Learning. Pour ce faire, nous intégrons les gradients physiques (dérivés des sorties par rapport aux entrées) aux modèles DL. Notre stratégie a donné des bons résultats vers une meilleure généralisation des réseaux de DL et notre étude théorique est corroborée par une validation empirique incluant une étude d’ablation.

\end{foreignabstract}

   \chapter*{Context}

This work was developed in the context of the internship of the advanced master's program HPC-AI Mines Paristech. The internship took place at the startup Extrality, in Paris.

Extrality frees industrial design and enables manufacturers in aerospace, transport and energy to drastically reduce their time-to-market. Concretely Extrality's platform of AI-powered simulations allows engineers to assess the aerodynamics, fluids and thermics of a product in just a few seconds while keeping industry-critical physical accuracy (four pending patents).     

It is composed of 17 international-level experts combining Machine Learning and Physics knowledge, and passionate about mastering tomorrow’s complex industrial and environmental challenges. This work is part of the company's R\&D efforts towards an innovative product with a more generalizable DL framework to meet the requirements of real-world use cases, including aerospace, railway and automobile applications. 

\begin{figure}[htb]
\includegraphics[width=3 cm]{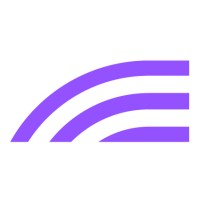}
\centering
\end{figure}
  \cleardoublepage
  \addcontentsline{toc}{chapter}{Acronyms}
  \printnoidxglossary[type=\acronymtype]
  \addcontentsline{toc}{chapter}{Variables}
  \printnoidxglossary[type=notation]
  \listoffigures
  \listoftables
  \tableofcontents

  \pagenumbering{arabic}
  \mainmatter
  \chapter{Introduction and motivation}
\label{Introduction}

Numerical simulations are of the utmost importance nowadays. They
allow projecting prototypes in a safe virtual environment with an affordable cost rather than actually building and testing them.
More specifically, \gls{CFD} simulations can be applied in diverse domains
such as aviation, automobile, oil \& gas, renewable energies,
etc.

However, the execution of these simulations relies on solving the complex \gls{NS} equations. To resolve these
\gls{PDE}, a computer can take many hours or days, depending on the physical domain
and discretization used to achieve the desired precision, even with the computational power provided by the advancement of
\gls{HPC} in the last decades.

In this context, the progress of AI, and more specifically of DL, has made possible the analysis of data from simulations and proposed new approaches on how to handle problems. Machine Learning presents a great capability to model physical problems when a possible input-output relationship is present, which is the case for \acrshort{CFD}. Furthermore, Deep Neural Networks enhance this capability by allowing almost any kind of non-linearity to be represented. Therefore, using DL can significantly help to speed up the solution of numerical simulations with reasonable precision.

Since Deep Learning models are statistical, it is well known that they are hungry; hence the more data is available, the better models are. As mentioned, the data is generated from expensive simulations, and new ways to improve generalization performance with the same amount of data remain an open problem with several research directions. One possibility, that is analyzed in this work, would be to supervise the network not only with the fluid dynamics, but also its derivatives, namely adjoint gradients.

The simplified \acrshort{NS} \acrshort{PDE} equations have the form:

\begin{equation}
\begin{aligned}
    \partial_t + u_xu + u_yv = - \frac{1}{\rho}p_x + v(u_{xx}+u_{yy}),  \\
    \partial_t + u_xu + u_yv = - \frac{1}{\rho}p_x + v(u_{xx}+u_{yy})
    \label{NvSkSimplified},
\end{aligned}
\end{equation}

Where \gls{$u$} is the velocity along x axis, \gls{$v$} the velocity in y, \gls{p} the pressure, \gls{rho} the density and \gls{nu} the viscosity.

 One way to determine the derivatives of the above system is using the Adjoint Method, which allows to compute the gradient of a specific function w.r.t. to its parameters when this function is constrained by a \acrshort{PDE}. This method presents itself as a fast and efficient alternative for this operation, regardless of such parameters. Hence, it is commonly used in the \acrshort{CFD} industry for shape optimization.

This work leverages the gradient data computed by the adjoint method in order to boost the generalization performance of currently used surrogate models in industry and academia. The framework presented can drastically reduce errors on unseen data, mitigating the need to acquire/generate extra simulations.

  \chapter{State of the art}
\label{stateoftheart}

In this chapter, research papers from different fields are concisely summarized. We make a literature review study and provide theoretical and benchmarking results as a core reference to make our thesis.

\section{Deep Learning in Computational Fluid Dynamics simulations}
In the context of Deep Learning for \acrshort{CFD} simulations, multiple models can be employed for different goals. The following variables can be predicted by the network:

\begin{itemize}

\item Volumetric field variables:

This is the most classical and intuitive application - to learn a model to predict the flow variables at each point of a structured grid. Here, \glspl{CNN} are widely used with great success, and a global overview of the simulation can be achieved, returning values such as pressure and velocity. Those models \cite{thuerey_deep_2020, um_solver---loop_2021, mohan_embedding_2020} are also commonly called \textit{voxelysed}, because they reconstruct the mesh to a voxelysed grid, allowing as well the treatment of surface variables. Appendix \ref{appenA} presents a finer explanation of such models.

\item Surface variables:

In the industry, computing field variables on the surface of the geometrical shape being simulated is vital. The stress and pressure values on the surface give an idea of how the shape reacts to the boundary conditions and how it can be changed/adapted to get to a better design. Treating directly the surface, however, is more complicated because of the unstructured form of the data and the physical phenomena at the boundary layer in fluid dynamics \cite{clauser_turbulent_1956}. Recently, new models appeared to deal with this data (meshes, graphs, points of cloud, manifolds, etc.), called \glspl{GDL}. Efficient and widely used models are PointNet \cite{qi_pointnet_2017}, PointNet++ \cite{qi_pointnet_2017-1} and \glspl{GNN} \cite{sanchez-lengeling_gentle_2021, zhou_graph_2021}. Appendix \ref{appenB} explains better how these geometric models work, while appendix \ref{appenC} addresses \acrshort{GNN}s in \acrshort{CFD}.

\item Performance variable 

Related to the previous point, instead of computing variables on the whole surface, a different model can be employed to calculate a single performance metric, which is usually the final goal of the simulation. Examples are the drag or lift coefficients of the design - corresponding, respectively, to the normalized forces along the x and y directions. They could also be computed by integrating the surface variables above, which is a better solution as it gives more generalizable results, considering the small nuances of the solution instead of a global final value. Hence, this approach was preferred for this work.

\end{itemize}

Modeling \acrshort{CFD} with DL requires a rigorous analysis of the results. As explained in appendix \ref{appenD}, not only quantitative, but also qualitative results need to be analyzed.

\section{Implicit and continuous neural representations}

To better model the relationship between input and output of the network, some feature transformations on geometric data and architecture modifications can be done. They come from the field of \gls{INR} \cite{mildenhall_nerf_2020, tancik_fourier_2020, sitzmann_implicit_2020, mehta_modulated_2021}.

These transformations are motivated by the fact that deep networks are biased towards learning lower frequency functions \cite{rahaman_spectral_2019}. Therefore, aiming to better represent higher frequency functions, which is the case of the surface fields we are interested in, these techniques can be useful. Appendix \ref{appenE} explains different alternatives in more details.

\section{Regularization through gradient}

In the last sections, it was explained how modeling \acrshort{CFD} simulations with Deep Learning yields good results and can be used to speed up these computationally expensive simulations. Nevertheless, the quality of the results depends directly on the training data available, and new techniques for improving generalization performance can help to achieve the same outcome fewer less observations at one's disposal.

One of those techniques, presented by \cite{huge_differential_2020}, is called Differential Machine Learning. The concept behind it is to use the gradients of the target labels w.r.t. the input data ($\frac{\partial y^i}{\partial x^i}$, where $\gls{x}^{\gls{i}}$ is the input and $\gls{y}^i$ the output) during training. It was proven that, in regression problems, this can largely improve performance results. The differential labels are used by the network and, as a consequence, accurate and fast functions can be learned from small datasets, decreasing also the need for regularization techniques.

This idea can be intuitive with simple examples: by learning not the points, but the slope of the function it is trying to be modeled, it should be easier for the algorithm to find the shape of this function (adjusted by the weights of the model). In fact, better results were found by supervising the network both with the points and the slope, i.e., $y^i$ and $\frac{\partial y_i}{\partial x_i}$.

Despite being applied for \acrshort{PDE}s finance (Financial Derivatives), it is believed that Differential Machine Learning could achieve great success also in a physics scenario, where the governing laws are also \acrshort{PDE}s. Especially in the case of \acrshort{CFD} simulations, where the functions to derive are smooth and the datasets are small and of high dimension, Differential Machine Leaning tends to work well. It's also important that gradients be of high quality.

\hfill \break
\noindent \textbf{Twin network}

\begin{figure}[htb]
\includegraphics[width=12cm]{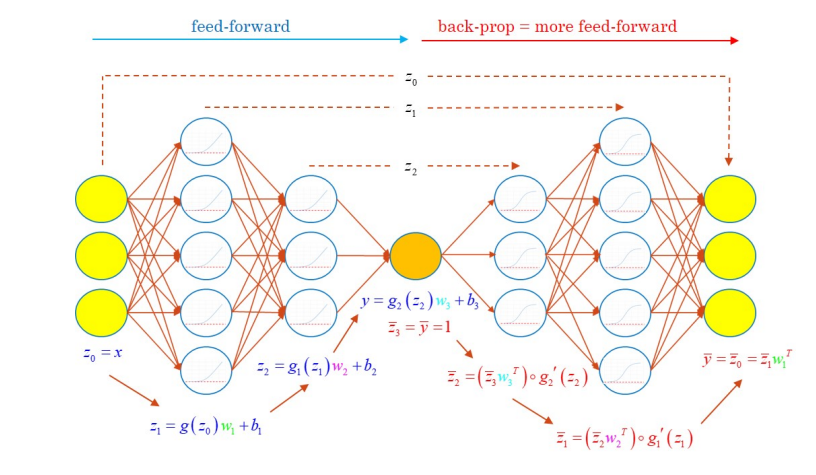}
\centering
\caption[Twin Network]
{Twin Network - the left side of the network is responsible for the feed-forward step to predict the target label. The right side mimics the architecture using shared weights, but back-propagates through the network to find the derivatives of the target labels w.r.t. the input \label{fig:twinNet}
\cite{huge_differential_2020}.}
\end{figure}

Differential Machine Learning penalizes the predictions to be constrained to the correct derivatives. Much like \acrshort{PINN}s \cite{raissi_physics-informed_2019}, it adds an extra term to the loss function during training. The architecture is called \textit{Twin Network} (figure \ref{fig:twinNet}) because the layers linked to one loss function are mirrored from the other one.

The model's gradient used during supervision is the same computed by \gls{AD} \cite{griewank_evaluating_2008} by the Deep Learning framework for backpropagation. In this way, the gradient loss is also backpropagated, making the model twice derived.

This back-propagation through the twin network is argued to require a continuous activation function. A normally classical choice, like \textit{ReLU} \cite{lecun_gradient-based_1998}, would not work, as its initial activation is not $\mathbb{C}^1$ and cannot be optimized. Suggested activation functions are: \textit{ELU} \cite{clevert_fast_2016}, \textit{Softmax} \cite{glorot_deep_2011} and \textit{SELU} \cite{klambauer_self-normalizing_2017}.

Differential Learning can also be interpreted as a kind of data augmentation technique. This is because extra information, besides the target labels, is used during training, having a regularization effect.

\section{Gradient in fluid dynamics: Adjoint method}

The differential labels explained above can be computed directly by \acrshort{CFD} programs. One efficient way to calculate the gradient of a \acrshort{PDE}, for example the \acrlong{NS} equations \ref{NvSkSimplified}, \textit{w.r.t} its input, for example the geometry, is the adjoint method. 

A simpler option to do this is by finite difference. It consists of perturbing the function \textit{w.r.t.} one of the input parameters and calculating the difference between states:

\begin{equation} \label{eq:finiteD}
    (\frac{\partial \bar{y}}{\partial x})_{m, n} = 
   \frac{\bar{y}_n(x + \epsilon e_m) - \bar{y}_n(x)}{\epsilon} + \mathbb{O}(\epsilon) \quad
   \cite{kenway_effective_2019}
\end{equation}

Where \gls{n} and \gls{m} are, respectively, the row and column indices of the matrix; \gls{epsilon} is the finite-difference step, and \gls{em} is a unit vector with a single unity in row \gls{m} and zeros in all other rows. This approach, although simple, has the disadvantage of calculating the \acrshort{PDE}, in our case the \acrshort{CFD} simulation, once for each parameter. If the number of parameters increases, the computational cost becomes prohibitive. This is exactly the case of fluid dynamics: the input is the position that, in the discretized approach, translates as each mesh cell/point position - thousands, or even millions.

Other methods, such as the complex-step or symbolic differentiation \cite{kenway_effective_2019}, present similar limitations. The adjoint method, on the other hand, makes it possible to compute a high-quality gradient that is independent of the number of parameters, being more efficient than the other approaches. For this reason, it is commonly used to find the Jacobian matrix in shape optimization tasks \cite{jameson_aerodynamic_2003}. Appendix \ref{appenF} explains in more detail the mathematics of the method and its two different techniques.

\section{Existing benchmaking datasets}

Although efficient surrogate models, like the ones mentioned in the previous section, have been developed to tackle physics problems, a standard benchmarking dataset is not available. The data becomes even scarcer when also the adjoint gradient is required.

Therefore, this work develops its own dataset, but based on previous papers. The objective is to have the most correct fluid simulations possible. A good alternative is to use the Turbulence Modeling Resource (TMR) of the Langley Research Center of the National Aeronautics and Space Administration (NASA) \cite{nasa_langley_research_cente_turbulence_nodate, ladson_effects_1988, ladson_pressure_1987, wadcock_structure_1979} on airfoils. The teams of the National Advisory Committee for Aeronautics (NACA) developed different airfoil families \cite{abbott_summary_1945}, which can be used in our application.

For simplification reasons, the focus will be only on the 4-digit family - which allows building a wide range of geometries by varying only 3 parameters that can be described in the following format: \textit{MPXX}. In this sequence, \textit{\gls{M}} is the maximum ordinate of the camber line, \textit{\gls{P}} is the position of this maximum from the leading edge, and \textit{\gls{XX}} is the maximum thickness.

A reliable, high-fidelity dataset constituting these conditions was developed by \cite{bonnet_airfrans_2022} and will serve as a reference for the primal equations (does not include the adjoint). The 2D airfoils presented are in the  steady-state subsonic regime at sea level and 298.15 K (25°C). In more detail, the flow presents a turbulent behavior, with Reynolds number between 2 and 6 million, is considered incompressible (Mach $\sim$ 0.3), and is  modeled by the \gls{RANS} equations with $k-\omega \: SST$ turbulence. The hexahedral mesh was created on Openfoam 2112 \cite{jasak_openfoam_2013} (the same application as the solver) based on the NASA developments for the NACA airfoils. Moreover, the mesh was developed to properly treat the boundary layer (\gls{y+} < 1 for all the design space).

  \chapter{Methodology}

This chapter describes the steps and procedures taken to get to the final framework, which trains the network with the gradient from CFD simulations.

\section{Synthetic dataset - polynomial}
\label{tests}

In practice, using the adjoint gradient during training for the NACA dataset is a complex task. To better understand the phenomena involved and develop a functional program, the approach was successfully tested on a synthetic case. For future reference, this experiment is entitled \textit{Task 1}. 

In this problem, data was generated based on a polynomial function having as input 1D points: $f(x) = y |   x, y \in$ $\mathbb{R}$ . The architecture was a shallow three layers \acrshort{MLP}, and the polynomial had degree 7.

This simple configuration was chosen to avoid saturating the data with the models developed, which would complicate the comparison between them. Moreover, the goal was to have an overfitting model, as in an industrial case, the difficulties are rarely the complexity of the architecture, but its generalization power. It is this characteristic we aim to test.

\begin{figure}[htb]
\includegraphics[width=12cm]{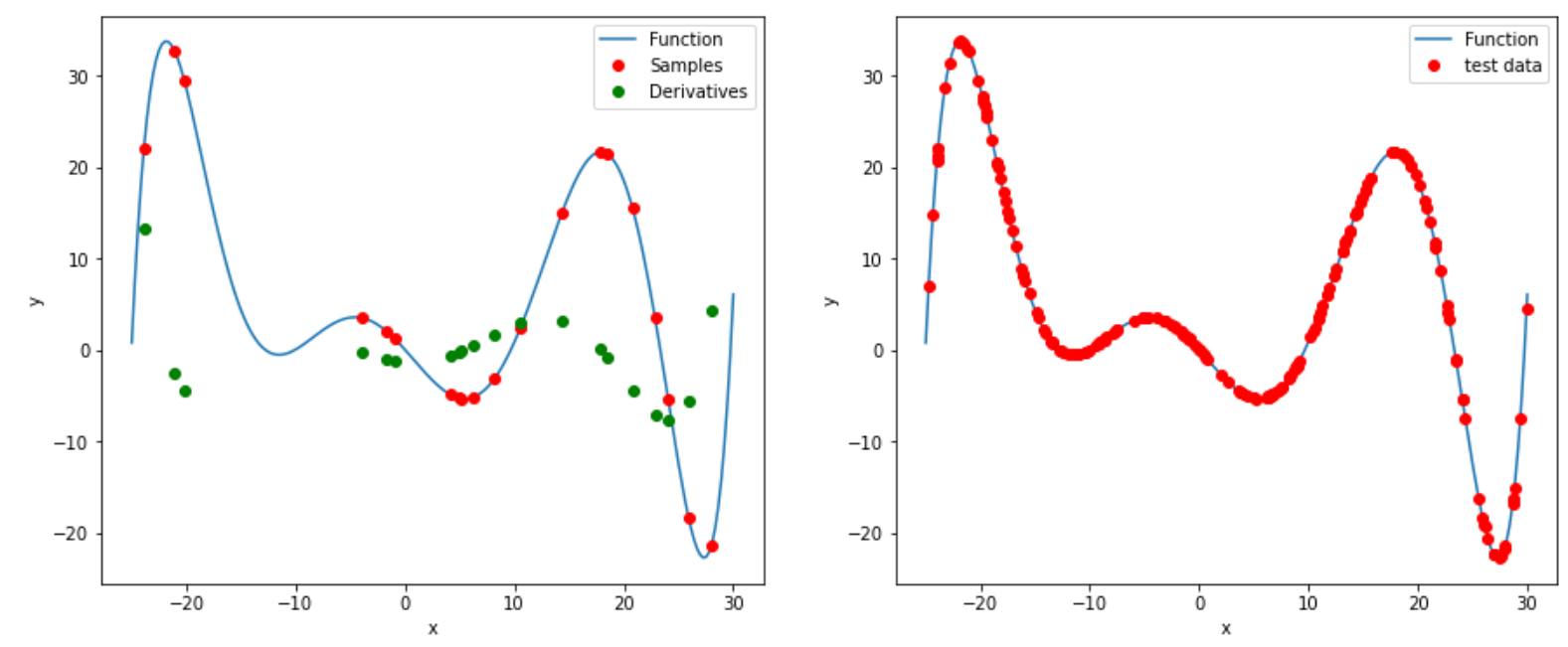}
\centering
\caption[Polynomial dataset sampling]
{Polynomial dataset sampling - on the left, the sample of training points and their derivatives. On the right, the testing points.}
\label{fig:SamplingPoly7}
\end{figure}

As the function is a polynomial, the gradient labels were analytically computed. Figure \ref{fig:SamplingPoly7} shows a sample of the dataset.

The objective was to compare the performance of the models using gradient/derivatives of the output w.r.t. 
the input ($\frac{\partial y_i}{\partial x_i}$) with the ones that only used the original labels $y$. Two gradient approaches were tested, as follows.

\subsubsection{Gradient models}
The first model was based on the idea of Multi-task Learning, where having a side, auxiliary task, can help the network in learning the main task. In this approach, it is used as target values of the network, not only the original data points $y_i$ but also their derivative labels. Adding this task, the model receives more information and would possibly predict better also the original target value, in a similar way to Multi-task Learning with shared parameters. For evaluation, however, only the original labels are used. This technique is referred to in this work as \textit{Both}, while the original case of using only the target labels is called \textit{Normal}.

The second option is Differential Machine Learning (\textit{DiffLearning}), already explained in chapter \ref{stateoftheart}. It is a similar idea, but here, the derivative ($\frac{\partial 
\bar{y}_i}{\partial x_i}$) is not passed as target value. but computed from the predictions ($\bar{y}_i$) with \acrlong{AD} and used to add an extra penalization term to the loss function, the difference
to the actual derivative points: $\frac{\partial \bar{y}_i}{\partial x_i} - \frac{\partial y_i}{\partial x_i}$. This approach is more computationally expensive than \textit{Both}.

\subsubsection{Data processing procedure}

For the model to learn correctly not only the target labels but also its derivatives, the gradient has to be processed. A necessary practice in deep learning is to scale the data - as much the output as the input.
However, by doing this, the gradient is also changed. Article \cite{huge_differential_2020} presents an essential way, thus, to also transform the derivative labels in the following manner:

\begin{equation*}
    \gls{haty}_i = \frac{\bar{y}_i - \gls{mu}_y}{\gls{sigma}_y}
    \quad\quad 
    \gls{hatx}_i = \frac{x_i - \mu_x}{\sigma_x}
\end{equation*}

\begin{equation}
     \frac{\partial \hat{y}_i}{\partial \hat{x}_i} = \frac{\sigma_x}{\bar{\sigma}_y} \frac{\partial \gls{bary}_i}{\partial x_i}
     \label{eq:scaling}
\end{equation}

This transformation will make the differential labels coherent with the gradient calculated by the framework (the inverse operation should be done to return to the true unscaled values). However, the resulting values themselves will not be scaled and could be much higher than the range [0,1]. To correct this issue, the following normalization can be done in the loss function:

\begin{equation}
    \gls{LG} = \frac{1}{N}\frac{\gamma}{||\frac{\partial \hat{y}}{\partial \hat{x}}||^2_b}\sum^N_{i=1}(\frac{\partial y_i}{\partial x_i} - \frac{\partial \hat{y}_i}{\partial \hat{x}_i})^2
    \label{eq:gradloss}
\end{equation}

$||\frac{\partial \hat{y}}{\partial \hat{x}}||^2_b$ can be computed before training for each batch. The hyperparameter \gls{gamma} is important to weight the loss of the gradient with the loss of the original labels.

A modification was made to equation \ref{eq:scaling}, since output data was scaled to the range [0,1]. In this way, the gradient should be scaled with the range of $\bar{y}$ instead of its standard deviation $\bar{\sigma}$.

\subsubsection{Model description}

An L2 loss (equation \ref{eq:normalloss}) is used for training and an L1 for testing. More information about the data generation setup and experimental results is given in chapter \ref{results}.

\begin{equation}
    \gls{LN} =  \frac{1}{N}\sum^N_{i=1}(y_i - \bar{y_i})^2
    \label{eq:normalloss}
\end{equation}

Considering \textit{DiffLearning}, one more choice is vital: the activation function. As mentioned in the article, the activation function should have special characteristics, excluding, for example, ReLU. Suggested alternatives are ELU, SELU and SoftMax. An uncited yet possibly efficient option would be \gls{SIREN} \cite{sitzmann_implicit_2020}. As explained in appendix \ref{appenE}, periodic activation functions have continuous and well-behaved derivatives, which could be an interesting characteristic to combine with \textit{DiffLearning}.

\begin{figure}[htb]
\includegraphics[width=12 cm]{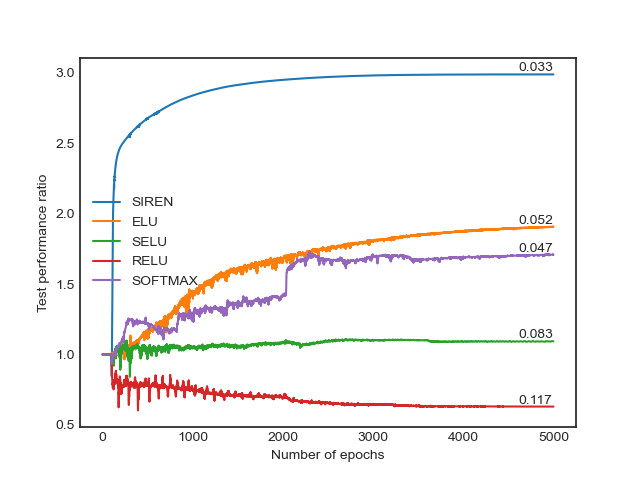}
\centering
\caption[Activation functions comparison for \textit{DiffLearning}]
{Activation functions comparison for \textit{DiffLearning} - ratio between the performance on testing of \textit{Normal} and \textit{DiffLearning} (final error written above each curve) for \textit{ELU}, \textit{Softmax}, \textit{\acrshort{SIREN}} and \textit{ReLU}  activation functions. Curves are averaged for 15 different sampling seeds, where the train set had 20 samples and the test had 200. SIREN stands out while ReLU doesn't work.}
\label{fig:Comparison}
\end{figure}

All the alternatives above were tested (figure \ref{fig:Comparison}). As expected, ReLU did not work. Among all the others, \acrshort{SIREN} had the best performance, which can be attributed to the characteristics mentioned. This activation is going to be used, thus, in all networks from this point on.

\subsubsection{Regularization effects}

The \textit{Both} approach did not yield gains in generalization performance. On the contrary, results on the test dataset were slightly worse than the \textit{Normal} case.

\begin{figure}[htb]
\includegraphics[width=15cm]{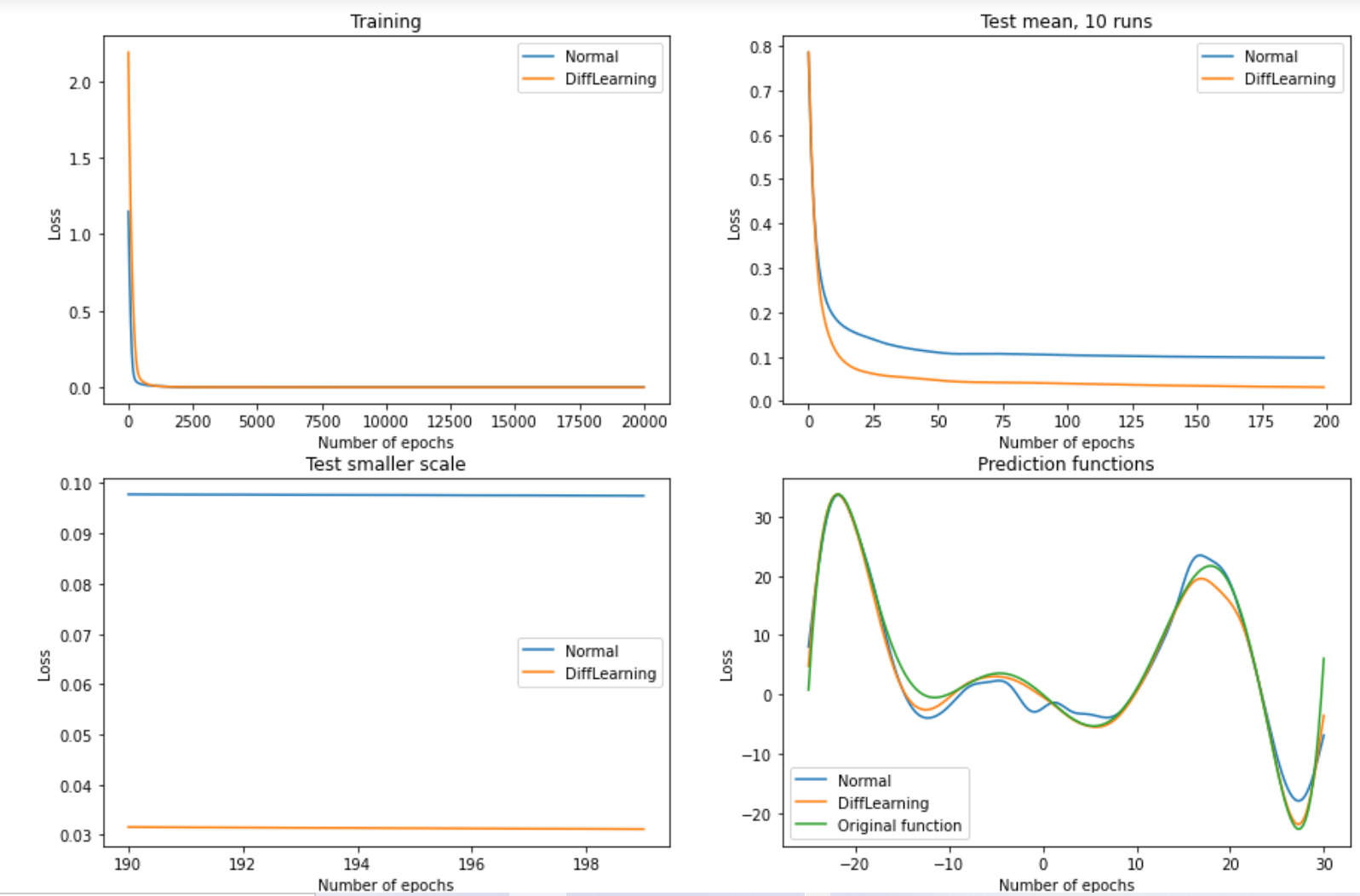}
\centering
\caption[Performance effects using gradient information]
{Performance effects using gradient information - performance on training (20 examples), testing (200 examples) and prediction.}
\label{fig:DiffLearningTraining}
\end{figure}

On the other hand, as it was expected and described in its paper, Differential Machine Learning works. It improves the performance on the test set, acting as a regularization technique to improve generalization performance. Both the final metric and the convergence are better than the \textit{Normal} case, as illustrated in figure \ref{fig:DiffLearningTraining}. More thorough experiments are explained in chapter \ref{results}.

\begin{figure}[htb]
\includegraphics[width=15cm]{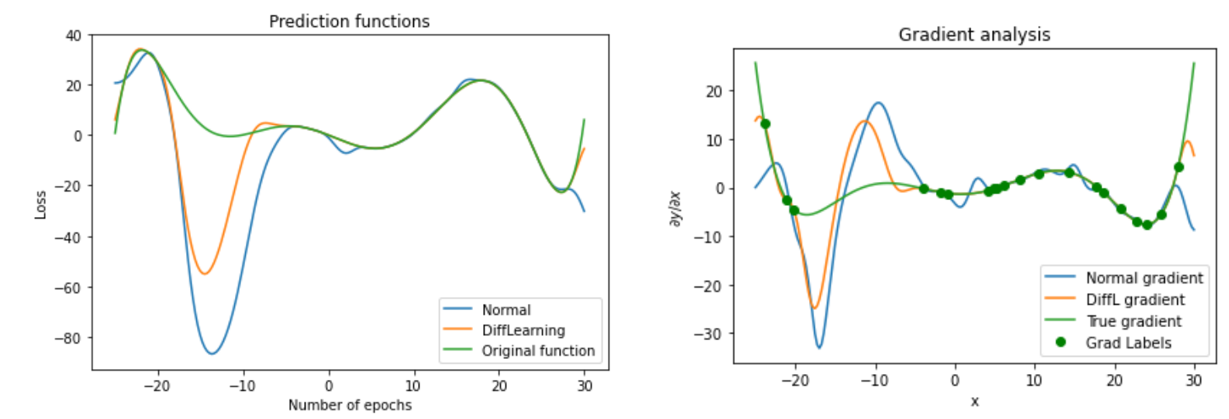}
\centering
\caption[Gradient analysis]
{Gradient analysis - on the left, predictions with and without \textit{DiffLearning}. On the right, the derivatives of those predictions w.r.t. the input. It can be noted that \textit{DiffLearning} respects more the true derivative function.}
\label{fig:GradientAnalysis}
\end{figure}

In addition, \textit{DiffLearning} yields a continuous model that respects the derivative of the true function (figure \ref{fig:GradientAnalysis}). This characteristic can be an advantage for the industrial case if the deep learning model is used for shape optimization.

\subsubsection{Improvements on Differential ML}

After proving the efficacy of the approach, two improvements were developed to enhance the architecture for the industrial case. The first one is to start training with only the \textit{Normal} loss and as of a certain epoch (a new hyperparameter of the model), switch for \textit{DiffLearning}. This is equivalent to adding the loss of the gradient (equation \ref{eq:gradloss}) to the normal one (equation \ref{eq:normalloss}).

\begin{figure}[ht!]
\includegraphics[width=15cm]{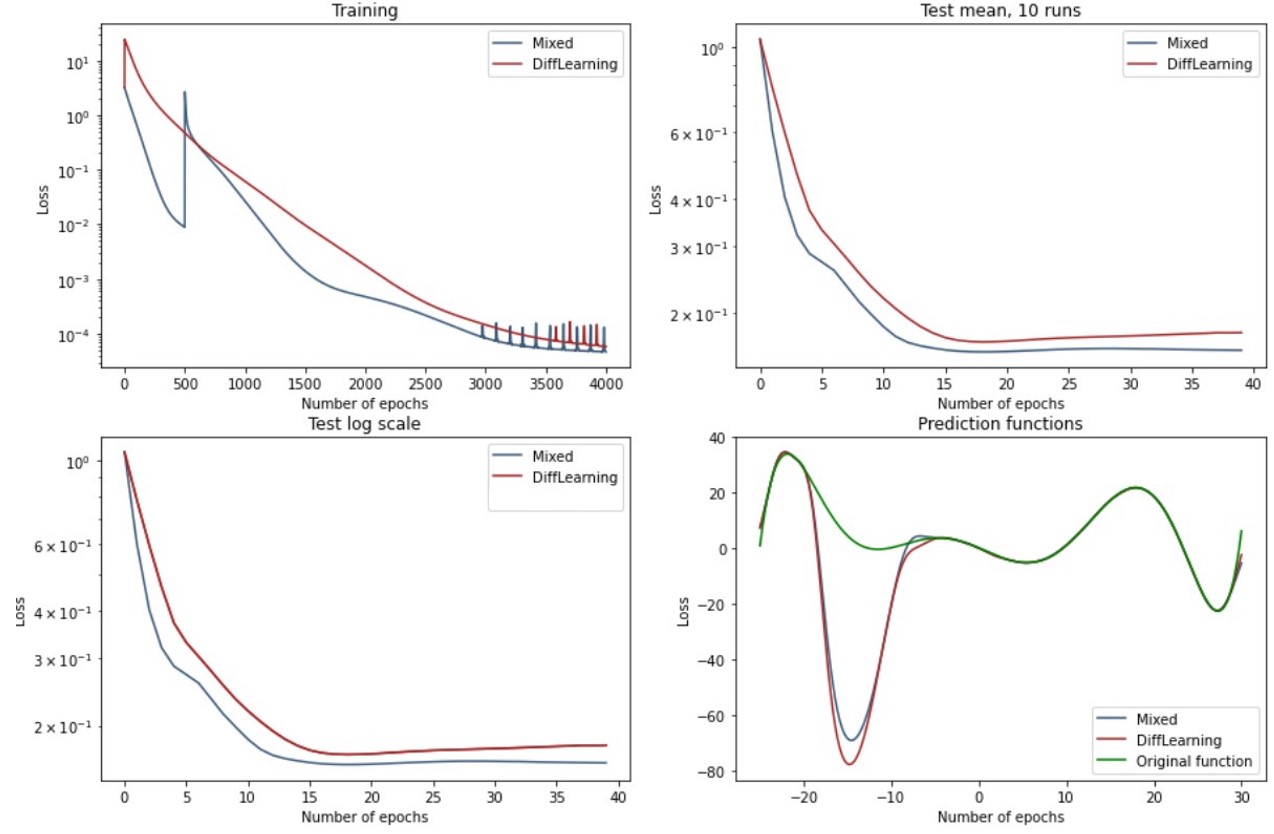}
\centering
\caption[Mixed training]
{Mixed training - train (20 samples), test (200 samples) and predictions for \textit{DiffLearning} model with (Mixed) and without 500 \textit{Normal} epochs.}
\label{fig:MixedTraining}
\end{figure}

This modification allows the network to converge to a more feasible solution before introducing gradient supervision. Otherwise, it can have difficulty at the beginning of the training with the reduced space of solutions, taking more epochs to descend. Figure \ref{fig:MixedTraining} shows how the training loss suddenly rises and testing loss descends when the gradient labels are taken into consideration.

The second modification was developed to solve a problem from the real NACA case. As the adjoint gradient is sometimes unstable and the CFD solver scale not always clear, a scaling parameter \gls{alpha} is added between the model's gradient and the gradient labels:

\begin{equation*}
    L_G = L2(\alpha \frac{\partial y}{\partial x}, \frac{\partial \hat{y}}{\partial x})
\end{equation*}

The parameter \gls{alpha} can be learned by the model and it was chosen to have its own optimizer with a specific learning rate. For testing, the analytical gradient was multiplied by a fake value. It was perceived that the optimization of \gls{alpha} was sensible to its initialization, to the true scaling it is trying to be achieved, and to the ratio between them. 



Begin training using exclusively the \textit{Normal} loss and then adding the gradient loss proved itself as a good convergence technique, especially for the alpha optimization. This strategy allows the model to acquire a feasible solution before introducing the gradient labels, as reducing the space of solutions in the early training can sometimes confuse the network.

To implement this along with the alpha optimization, the loss from the gradient has to be computed but detached from the model. In this way, when the backpropagation is done, the model's parameters are not influenced by the gradient loss, only \gls{alpha}.

\section{Dataset generation}
\label{Dataset}

As previously stated, the NACA \acrshort{CFD} dataset is developed in Openfoam. The reason is to profit from \cite{bonnet_airfrans_2022} setup for the primal solver while having an open-source and high-quality tool. Two obvious choices arise for solving the adjoint equations:

\begin{itemize}
    \item adjointOptimisationFoam - the own Openfoam's solution for the continuous adjoint. It has the analytical implementations for Spalart-Almaras and K-omega SST (only in version 2206) models.
    
    \item DASimpleFoam - a solver from DAFoam for incompressible flow using the discrete adjoint and \acrlong{AD}. DAFoam disposes of many solvers adapted to the original Openfoam solvers, making it possible to derive various models.
\end{itemize}

Despite the advantages of the continuous approach mentioned in F, the discrete option seems the best one for this application. It computes the exact gradient from the primal solution, being a high-quality data to use during training. The continuous gradient,
on the other hand, will only be exact in the limit of an infinitely fine mesh.

Nevertheless, as a first and simpler implementation, this work decided to begin with adjointOptimisationFoam. In terms of design space, two datasets were developed. The first one included only shape variation, and thus, the freestream velocity \gls{U} and \gls{aoa} are maintained constant, as shown in table \ref{table:DesignSpace}.

\begin{table}[H]
    \center

\centering
\begin{tabular}{ccc|cc}
  \hline
  \multicolumn{3}{c|}{4-digits} &
  \multicolumn{2}{c}{Flow variables}  \\
  \hline
   M & P & XX & Mach & AoA \\
  \hline
  [0, 6] & [0, 6] & [5, 25] & 0.2 & 2 \\
  \hline

\end{tabular}
    \caption[NACA simulations geometry design space]
    {NACA simulations geometry design space - where M is the maximum ordinate of the camber line, P is the position of this maximum from the leading edge, and XX is the maximum thickness of the wing. The Mach represents the freestream velocity w.r.t. the speed sound with the angle of incidence AoA.}
    \label{table:DesignSpace}
\end{table}

The second one did the opposite: varied the boundary conditions while using the same airfoil, a NACA 0012, as exemplified in table \ref{table:DesignSpace2}. The goal of generating both datasets is to isolate the design space seen by the model at each time, testing if Differential Machine Learning is useful even though the gradient is only w.r.t the position. 

\begin{table}[H]
    \center

\centering
\begin{tabular}{ccc|cc}
  \hline
  \multicolumn{3}{c|}{4-digits} &
  \multicolumn{2}{c}{Flow variables}  \\
  \hline
   M & P & XX & Mach & AoA \\
  \hline
  0 & 0 & 12 & [0.1, 0.25] & [-2, 5] \\
  \hline

\end{tabular}
    \caption[NACA simulations boundary conditions design space]
    {NACA simulations boundary conditions design space - where M is the maximum ordinate of the camber line, P is the position of this maximum from the leading edge, and XX is the maximum thickness of the wing. The Mach represents the freestream velocity w.r.t. the speed sound with the angle of incidence AoA.}
    \label{table:DesignSpace2}
\end{table}

\noindent\textbf{Continuous adjoint dataset}

The first choice to make is between the two available turbulence models: Spalart-Almaras \cite{kostic_review_2015} and  $K-\omega \: SST$ \cite{menter_ten_2003}. Both of them were verified. The first one, despite being simpler, exhibited a lot of instability issues, noisy results and extreme values.

Those problems, even if still present, were alleviated in the $K-\omega \: SST$ implementation. In addition, the choice of this model allowed us to follow a similar setup to \cite{bonnet_airfrans_2022} - being, thus, the preferred option. A common issue when deriving the equations in question is to consider that the turbulent viscosity field does not change with the shape during the optimization (\textit{Frozen Turbulence}), which can lead to incoherent sensitivity maps. The current implementation, however, includes the option of \textit{differentiated turbulence}, chosen for the simulations.

Most of the primal simulation parameters were inspired by \cite{bonnet_airfrans_2022}. For example the meshing strategy, a linear upwind discretization scheme and the boundary conditions for each field. The adjoint boundary conditions, on the other hand, a sensitive topic for the continuous adjoint approach, followed the official Openfoam's documentation suggestions and NACA tutorials.

Under-relaxation factors were used and tuned for each field to find the best relationship between the stability of the solution and convergence speed. In this context, an important parameter to improve convergence of the adjoint equations is the Adjoint Transpose
Convection (ATC) model. The ATC term, present in the adjoint momentum equations, can cause convergence difficulties \cite{kavvadias_proper_2015, skamagkis_eciency_2020}. Thus, a smoothing strategy that adds and subtracts this term multiple times during computation, was used. The domain was divided into 16 parts for a faster parallel run.

\begin{wrapfigure}{l}{8cm}
\includegraphics[width=8cm]{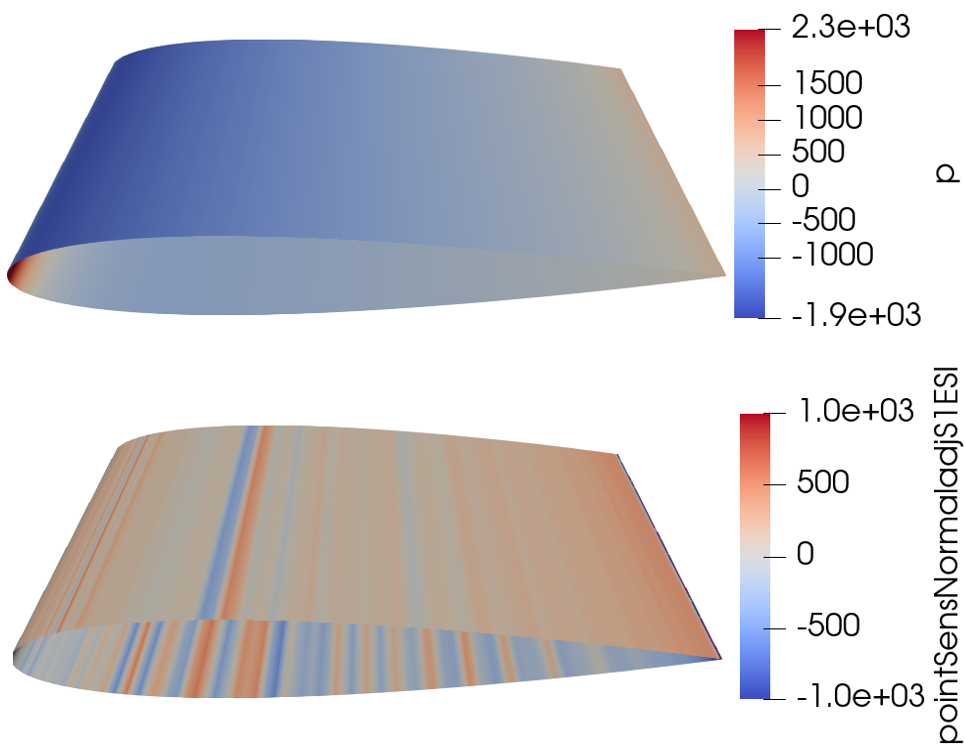}
\centering
\caption[Coarse mesh simulation]
   {Coarse mesh simulation - NACA 0012 $K-\omega \: SST$ continuous adjoint simulation on Paraview \cite{ayachit_paraview_2015}; a coarse mesh. On top, the heatmap of the pressure (Pa) distribution on the faces (5500 iterations). On the bottom, the heatmap sensitivity projected in the normal (extra 14300 iterations), using $\pm$ 1e+3 custom range.}
\label{fig:CoarseSimulation}
\end{wrapfigure}

Another important factor is the choice of the objective function and design variables used in the derivation. As mentioned in chapter \ref{stateoftheart}, in industry we are often interested in predicting performance variables, such as the drag and the lift - the forces respectively along the x and y axis, projected to the system rotated by the \acrlong{aoa}. Those values can be computed per cell by the solver by subtracting the \textit{wall shear stress} from the \textit{pressure} in the surface. Finally, the lift was selected as target, because its sensitivity presented a more informative distribution for the NACA wings than the drag.

Concerning the derivation, the recommendation is to use the nodes instead of the faces. When computing sensitivity maps, the variation in the normal vector is better posed when differentiating w.r.t. points. Therefore, the objective is derived w.r.t. the normal displacement of boundary points. In addition, the area was included in the final value.

With the setup ready, the residuals of the field variables, primal and adjoint, were analyzed to ensure convergence, as well as the total lift value. Nevertheless, even after careful choice of parameters 
and boundary conditions, the adjoint sensitivity presented a unstable behavior (figure \ref{fig:CoarseSimulation}) with extreme values on the trailing edge. These extreme values, considered non-physical outliers, made training deep learning models hard, and should be avoided.

\begin{wrapfigure}{l}{8cm}
    \includegraphics[width=8cm]{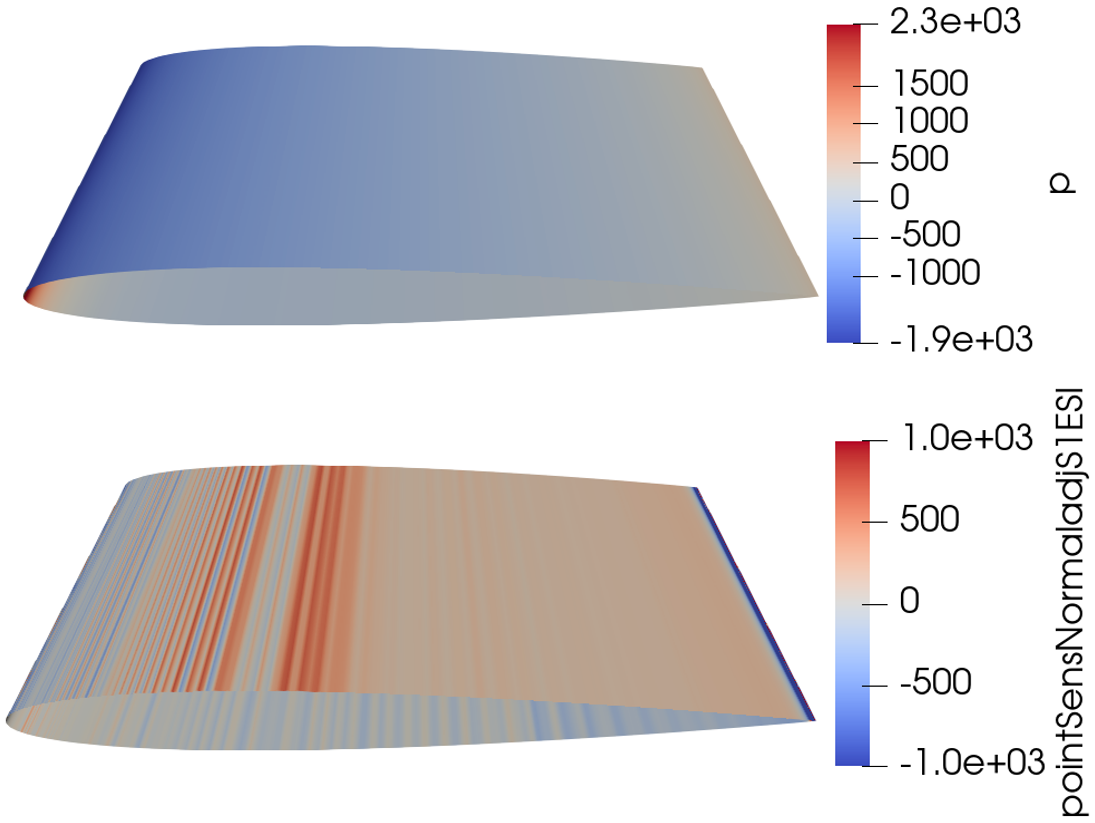}
    \centering
    \caption[Final simulation]
    {Final simulation - NACA 0012 $K-\omega \: SST$ continuous adjoint simulation on Paraview; a fine mesh. On top, the heatmap of the pressure (Pa) distribution on the faces (5500 iterations). On the bottom, the heatmap sensitivity projected in the normal (extra 14300 iterations), using $\pm$ 1e+3 custom range.}
    \label{fig:FinalSimulation}
\end{wrapfigure}

These characteristics can be associated with the continuous approach, where a coarse mesh and not fully converged 
primal can result in imprecise gradients. Those difficulties were damped with a mesh refinement and iterations convergence study, which made the simulations much heavier, but more reliable, as shown in figure \ref{fig:FinalSimulation}. It should be noted when analyzing both simulations that a finer mesh with the same number of iterations results in a less converged primal and a less precise adjoint than it could be. 

The amount of iterations was maintained constant among simulations, using a number that assured convergence for them all while having a feasible computational cost. For the NACA 0012 example (figure \ref{fig:FinalSimulation}), the final airfoil mesh had \textbf{2560 nodes} and \textbf{1580 cells}, as the 2D simulation requires a 3D representation. If considering the whole domain, the number of elements is much more elevated.

\section{Fake target NACA dataset}

There is a big gap between the polynomial case (Task 1) and the real NACA dataset. The input, for example, is a multi-dimensional geometrical data, and the model should come from \acrshort{GDL}. The gradient, in its turn, is no longer analytical, but generated by the adjoint method, being noisy and unstable. In order to test the first two points while maintaining an analytical gradient, this test was developed. For future reference, this experiment is entitled \textit{Task 2}. 

The input data was chosen to be exactly the same. As mentioned in section \ref{Dataset}, the target to predict is the lift force, which in practice can be computed by integrating the forces on the surface of the airfoil. The entry data, thus, should describe the surface of the wing. As further discussed in chapter \ref{stateoftheart}, normally this is represented by a Point Cloud, where the coordinates x and y can be used to describe the geometry. In addition, as NACA wings present a sharp shape at the edges, its description can be improved by using also the normals ($n_x$, $n_y$) to the 2D surface. In this way, the features have 4 dimensions.

To represent this unstructured geometric data, from the options mentioned in appendix \ref{appenB}, PointNet was chosen for being a simple and widely used architecture for berchmarking in computer vision. The output, in its turn, is generated by a 4th-degree polynomial function: $y=f(x, y, n_x, n_y)$. In this way, the gradient labels can be analytically sampled.

\begin{figure}[htb]
\includegraphics[width=10cm]{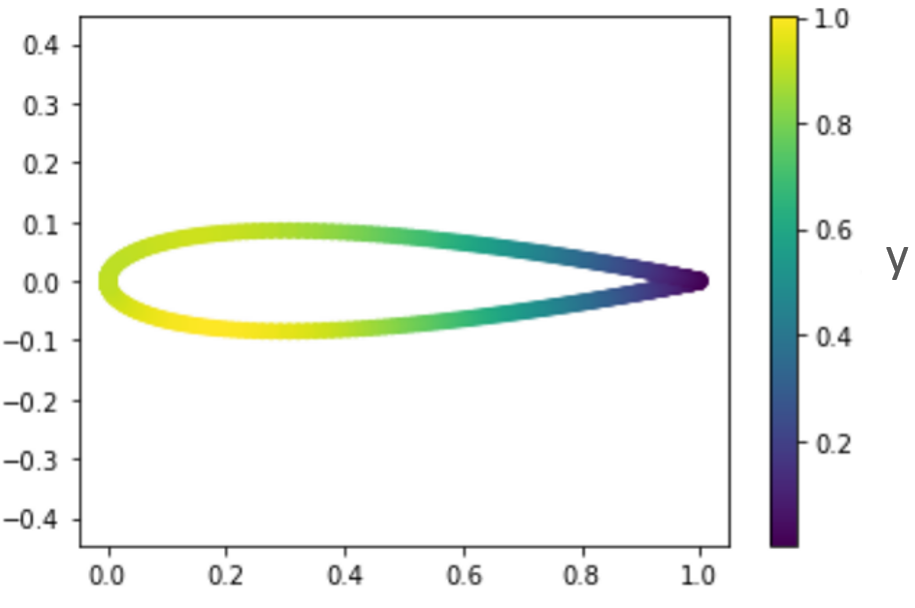}
\centering
\caption[Fake target NACA example]
{Fake target NACA example - Heatmap of the fake dataset. The example shows a NACA 0012, whose coordinates are used by the polynomial function to output a value \textit{y} per point.}

\end{figure}

The input and output data are scaled. The normals are already normalized and dispense further treatment. The coordinates follow the recommendations of the PointNet paper and are divided by the radius of the encompassing sphere (circle in this case). The $\bar{y}$ target, in its turn, suffers the scaling of the flattened PointCloud, containing all the values from all the simulations. The same scaling steps as in the polynomial (equations \ref{eq:scaling} and \ref{eq:gradloss}) are applied for the gradient, adapting the formula. 

The gradient itself should also be treated differently. The ajoint generated in section \ref{Dataset} is projected in the normal direction. Even though Openfoam gives the option to compute a vectorized, unprojected value, this is the most common form. Therefore, the first two dimensions of the model's gradient, representing the position, are projected in the normal and used during training.

The results in the test set were also promising, even tough, the task was too easy for the network and the potential of the approach could not be fully explored.

\section{NACA dataset task}

The two generated datasets (\ref{table:DesignSpace}, \ref{table:DesignSpace2}) were analyzed. For future reference, these experiments are entitled, respectively \textit{Task 3} and \textit{Task 4}.

The methodology applied to Task 3 mimics the implementation of Task 2 on what concerns the input and the model. The same features with the same preprocessing were used.

For Task 4, however, as there's no geometry variation, it is useless to use a PointNet. A simple dense network (MLP) with periodic activation functions (SIREN) suffice. Thus, each point continues to be treated separately, but encode as features the velocity in x and y (similarly to \cite{thuerey_deep_2020}) as well as the position.

The output, in its turn, is the same for both - data generated from the simulations, i.e., the lift following the \acrshort{aoa} direction and its adjoint derivative projected on the normal.

\subsection{Description of the model's target}

As stated in chapter \ref{stateoftheart}, predicting surface variables instead of the total performance metric helps the network to better represent the function. Thus, the lift per point was chosen to be the prediction target.

Another option would be the lift per cell. Nevertheless, as the derivation was chosen to be done w.r.t. the points, in order for the model to have a coherent gradient, an interpolation of one of those variables, lift or sensitivity, would have to be made. The interpolation points-cell, however, is not recommended because Openfoam computes variables in its cells' centers and interpolates them to the nodes; which would result in a double interpolation in the above case.

Moreover, the punctual lift is already multiplied by the correspondent area, so the derivation corresponds to the ground truth sensitivity.

\subsection{Preprocessing}

\begin{wrapfigure}{l}{8cm}
\includegraphics[width=8cm]{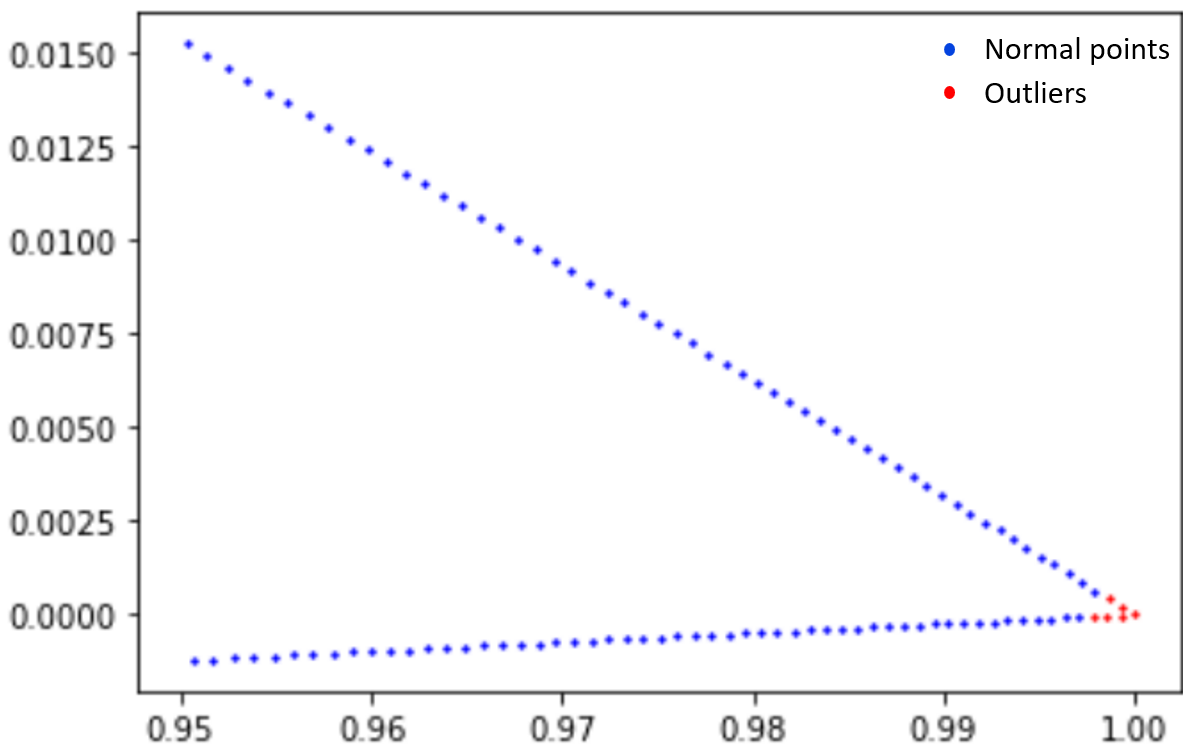}
\centering
\caption[Gradient outliers in NACA wings]
{Gradient outliers in NACA wings - outliers identified with a Z-score greater than 3. A zoom from a NACA wing's trailing edge, the regular points in blue and the outliers in red.}
\label{fig:Outliers}
\end{wrapfigure}

As it was exposed, the continuous adjoint gradient is unstable and presents outliers on the trailing edge, where the shape is sharper. One way to identify outliers is by calculating a Z-score \cite{abdi_z-scores_2007}. Image \ref{fig:Outliers} illustrates the typical behavior in this region.

In order to treat this, different alternatives arise:
\begin{itemize}
    \item Limit the gradient to a certain value per geometry - an option to implement this would be to use the Z-score with 3 standard deviations as limit, correcting only the outliers.
    \item Gaussian filter - preprocessing the derivative labels with a filter has the advantage of smoothing its distribution and damping extreme values.
    \item Remove the outliers for gradient training - different criteria to implement this can be used. 
\end{itemize}

Usually, the three alternatives could be separately used or combined. However, it was noted that the values on the trailing edge were not only extreme but also non-physical and oscillatory. Therefore, limiting them did not work as it resulted in an incorrect supervision. Applying simply a Gaussian filter proved to be faulty, as the nonphysical information was propagated. Finally, a good option was to remove the gradient values on the trailing edge during training. Instead of doing a Z-score identification, it was chosen to do a removal by coordinate ($x>x_{limit}$), in order to ignore flawed points that were not outliers.

In this way, a Gaussian filter could also be applied without propagating the wrong information. In some cases, smoothing the derivatives worked well.

Even with this preprocessing, the magnitude of the gradient was much bigger than the actual targets. This fact made training really hard for the network, despite the loss normalization (in equation \ref{eq:gradloss}). Accordingly, the network learning was facilitated by the common technique of scaling the output. An appropriate method has to be applied since the standard scaling could change the sign of the derivatives and, as a consequence, its physical meaning. A proper solution encountered was to divide by the range:

\begin{equation}
    \frac{\partial \hat{y}}{\partial \hat{x}} =
    \frac{\partial \bar{y}}{\partial x} /
    \delta
    \label{eq:scalingAdj}
\end{equation}

Where $\delta$ represents the range (difference between the maximum and the minimum) of the gradient labels. This operation makes the other scaling techniques in equations \ref{eq:scaling} and \ref{eq:gradloss}, used in the synthetic datasets, unnecessary.

\subsection{Description of the scaling parameter}

Although necessary, applying the filter and scaling the gradient label will make it incoherent with the one from the model. In this way, the use of the scaling parameter \gls{alpha} introduced in section \ref{tests}, becomes fundamental.

The theoretical optimum value of \gls{alpha} could be found by using the range of equation \ref{eq:scalingAdj} with equation \ref{eq:scaling}. In practice, nonetheless, it was noted that the true value was different, probably because of Openfoam's hidden units and the adjoint instability. The \gls{alpha} optimization has also proven itself to be a difficult task. To understand the reasons, we came back to the polynomial case.

\begin{figure}[htb]
\includegraphics[width=\textwidth]{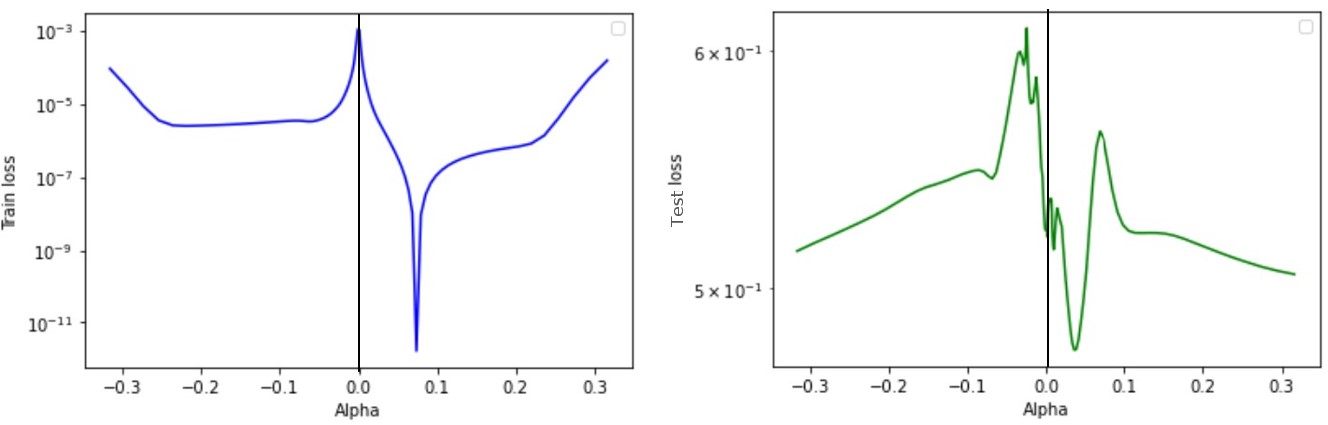}
\centering
\caption[\gls{alpha} curves for 5 samples on training set]
{\gls{alpha} curves for 5 samples on training set - performance curves for 5 training samples (left) and 200 for testing (right) for different values of \gls{alpha} (fixed) when the ground truth was a 0.1 factor.
\label{fig:Alpha5samples}}
\end{figure}

In fact, when little information about the gradient is available, i.e., scarce training examples or imprecise derivatives, the model has trouble optimizing \gls{alpha} and finding its real value. Figure \ref{fig:Alpha5samples}
plots the model's performance for different values of \gls{alpha} and shows how the minimum does not match the ground truth. In addition, positive and negative values present a certain symmetry around 0, which becomes more important when less gradient information is available to the network - fewer samples or projection on the normal (verified in Task 2). 

\begin{figure}[t]
\includegraphics[width=\textwidth]{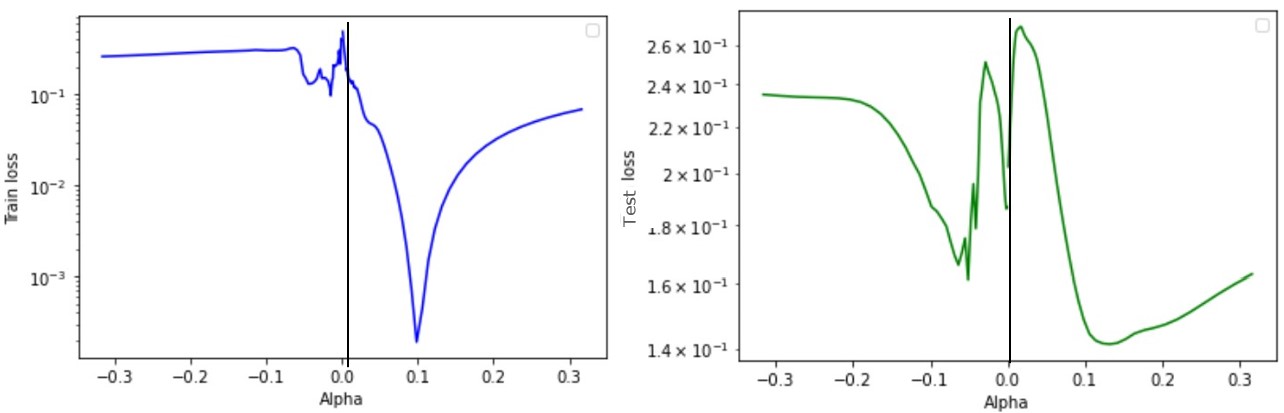}
\centering
\caption[\gls{alpha} curves for 20 samples on training set]
{\gls{alpha} curves for 20 samples on training set - reproducing figure \ref{fig:Alpha5samples} with more training examples.}
\label{fig:Alpha20samples}
\end{figure}

Figure \ref{fig:Alpha20samples}, on the other hand, illustrates how the symmetry disappears and the optimization works with enough training samples. Those conclusions were extended to the real NACA case, where symmetry was observed and the optimization was difficult. The model viewed all available data in order to plot the same curves and infer the value of \gls{alpha} closer to the truth. Even though the symmetry phenomenon remained, the chosen digit was fixated and yielded good results.

  \chapter{Experimental results}
\label{results}

\section{Software and libraries}

In order to build on the benchmark \cite{bonnet_airfrans_2022}, Openfoam was chosen as a primal solver, generating vtk result files. Version 2206 had to be used to benefit from the most current methods concerning the adjoint. For the adjoint itself, Openfoam's continuous solver was used. The processing power for those parallel simulations was 16 \textit{n1-standard GCP} cores and 60 Go of memory.

The deep learning framework is \textit{pytorch} \cite{paszke_pytorch_2019}, which uses \acrshort{AD} for computing the gradient to be backpropagated during training and can be leveraged for Differential Machine Learning - a tool called \textit{autograd}. In addition, \textit{pytorch geometric} \cite{fey_fast_2019} is used to build graph architectures, and \textit{pyvista} \cite{sullivan_pyvista_2019} to treat the vtk data from the simulations. During training, an \textit{NVIDIA Tesla P100} GPU was used, disposing of the same memory as in the simulations.

\section{Comparative study}

The performance of the Differential Learning framework is compared with the simple baseline, which uses only the loss considering the original target labels and is referred here as \textit{Normal}. An interesting piece of information to extract is how both methods compare when different amounts of data are available. 

In this way, parting from the principle that, in theory, one complete adjoint simulation has the cost of two primal simulations, it can be concluded if it is worth it or not to generate adjoint data instead of only the primal. Notwithstanding, even if the framework does not match the performance with half the data, adjoint simulations are already done in industry for shape optimization purposes, and retrieving this information would be advantageous.

Having that in mind, an interesting comparison can be done between the learning curves of both approaches - measuring the performance for different train set sizes. This study can be done for each of the three benchmarks: the polynomial (Task 1), the fake target NACA (Task 2) and the NACA datasets (Task 3 and 4).

\subsection{Model description}

The optimization of all models was done with ADAM optimizer, using an adaptive exponential learning rate scheme. On tasks 1, 3 and 4, the hypermarameters were optimized by grid search for each model (different architectures and training sizes). A better option still, would be to use a statistic hyperparameter optimization \cite{turner_bayesian_2021}, for example Bayesian, which yields better results. Task 2 presented a debugging purpose and, despite the good outcome, no further optimization was done with it.

The networks size, in its turn, was kept constant for each benchmark. Table \ref{table:modelSizes} shows the sizes of those models.

\begin{table}[H]
    \center

\centering
\begin{tabular}{c||cccc}
    \hline
   \textbf{Benchmark} & Task 1 & Task 2 & Task 3 & Task 4\\

  \textbf{Backbone} & \acrshort{MLP} & PointNet & PointNet & MLP\\

  \textbf{Parameters} & $5\cdot10^1$ & $5\cdot10^4$ & $3\cdot10^6$ & $1\cdot10^4$\\
  \hline

\end{tabular}
    \caption[Model's specification]
    {Model's specification - the 4 models used for the 4 datasets evaluated. As the complexity of the task increased, the complexity of the model followed it, having more parameters.}
    \label{table:modelSizes}
\end{table}

\subsection{Visualization: learning curves}

For all benchmarks, the training loss is an L2 between the predictions and ground truth solution in each point. For testing, it is different. An L1 is used in Tasks 1 and 2 between the predictions and target in each point. In Task 3 and 4, however, the total lift (summation of the punctual lifts) was chosen and inserted in a mean relative error loss. This metric is more interesting for a real use case as it indicates the total performance of the design.

The entire datasets were devided in training and testing. For Tasks 1, 3 and 4 an extra validation set was used with an L1 loss per point (only for NACAs). 

In practice, while the test examples were maintained fixed during different runs to allow comparison, the train set changed. The reason is that predictions varied considerably in different runs. \textit{Pytorch geometric}, for once, has irreproducible results on GPU, which, especially in the synthetic case, changed with the parameters initialization seed. The behavior was more stable in bigger models. Notwithstanding, the sampling seed also had an important influence in the results and comparisons.





To get trustworthy conclusions, a statistic strategy was implemented, running several experiments with different seeds (sampling and initialization). The minimal error was taken for each run instead of the last one, as, in practice, the error rises and an early stop proceeding could be used. 

The number of times each training was run depended of the confidence wanted. A p-value of 0.05, representing a 95\% confidence, was chosen as threshold. The null hypothesis ($H_0$) is that the error from the \textit{DiffLearning} model $e_D$ is the same from the \textit{Normal} one $e_N$, i.e., $\mathbf{E}(e_D - e_N) = \mathbf{E}(e_D) - \mathbf{E}(e_N) = 0$. 

\begin{wrapfigure}[18]{l}{8cm}
    \includegraphics[width=8cm]{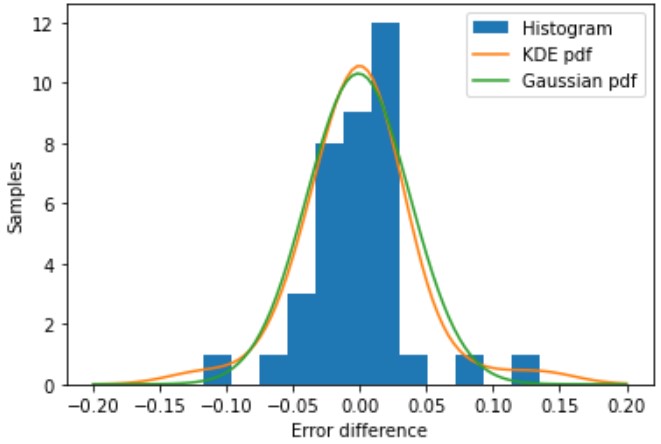}
    \centering
    \caption[Performance distribution analysis]
    {Performance distribution analysis - histogram of the difference between \textit{DiffLearning} and \textit{Normal} models on a 10 samples training set (task 4). The sample was fitted by a Kernel Density Estimation and a Gaussian distribution with the same mean and variance. As it can be note, both have similar shapes, indicating that the sample is normally distributed.}
    \label{fig:KDE}
\end{wrapfigure}

Rejecting this hypothesis means that they are not the same and \textit{DiffLearning} has a lower error (or the inverse). Defining the variable $e = e_D-e_N$, we can apply a T-student test to verify the null hypothesis for the sampling of $e$ computed for each dataset. The t-score assumes the following formula:

\begin{equation*}
    T = \frac{0 - (\mu_D - \mu_N)}{\sqrt{\frac{\mathbf{V}_D + \mathbf{V}_N}{N}}}
\end{equation*}

Where $\mu$ and $\mathbf{V}$ are, respectively, the mean and variance of the sampling of $e$ of size $N$. This equation can only be applied under the formulation that $e_D$ and $e_N$ are independent variables and $e$ follows a Gaussian distribution. The first one is a reasonable hypothesis, while the second can be verified by a Kernel Density Estimation \cite{chen_tutorial_2017} (figure \ref{fig:KDE}).

As it is shown in the next section, not all the results achieved a p-value permitting definitive conclusions, despite the large number of experiments.

\subsubsection{Task 1}

\begin{figure}[htb]
    \includegraphics[width=13cm]{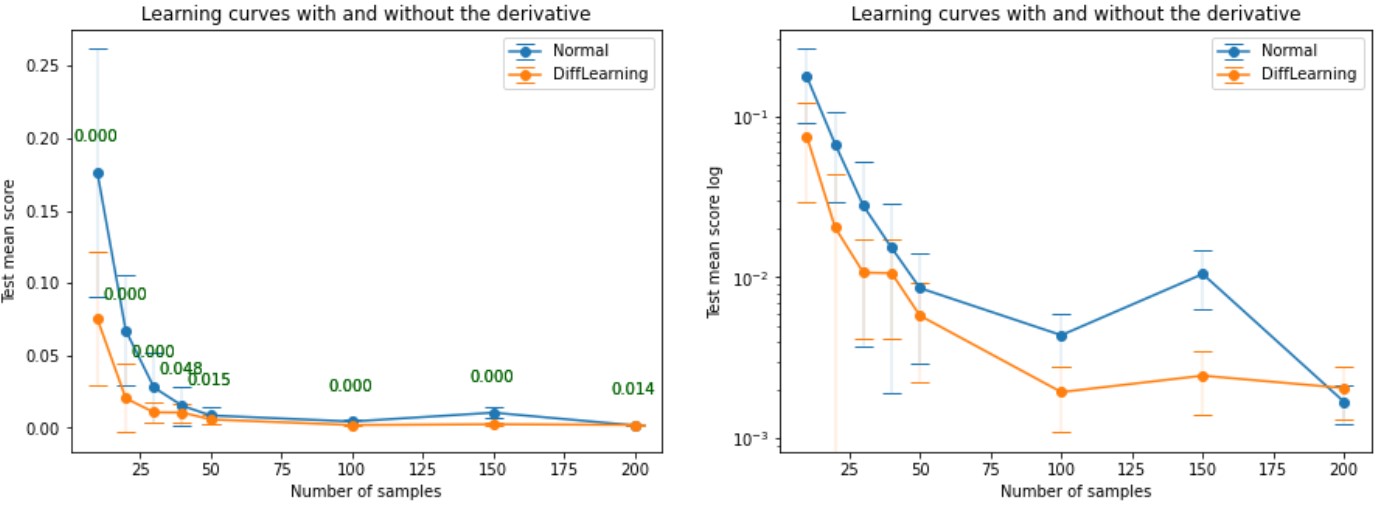}
    \centering
    \caption[Differential Machine Learning on polynomial]
    {Differential Machine Learning on polynomial - comparison between Differential Machine Learning and \textit{Normal} approaches for Task 1 using different sample sizes for training. The graphic on the right is in log scale. Green and red annotation represent the p-value that, respectively, respect or not the threshold determining the equality of the models}

    \label{fig:FinalPolynomialDiff}
\end{figure}

In the polynomial dataset, as the number of samplings was unlimited, 200 testing samples were generated to cover all the design space (figure  \ref{fig:SamplingPoly7}) while 50 served for validation. Training samples were, thus, generated independently according to the need. Figure \ref{fig:FinalPolynomialDiff} shows the learning curves for Task 1, including the standard deviation of the mean values. It is noticeable the potential of Differential Machine Learning.

\begin{figure}[htb]
    \includegraphics[width=13cm]{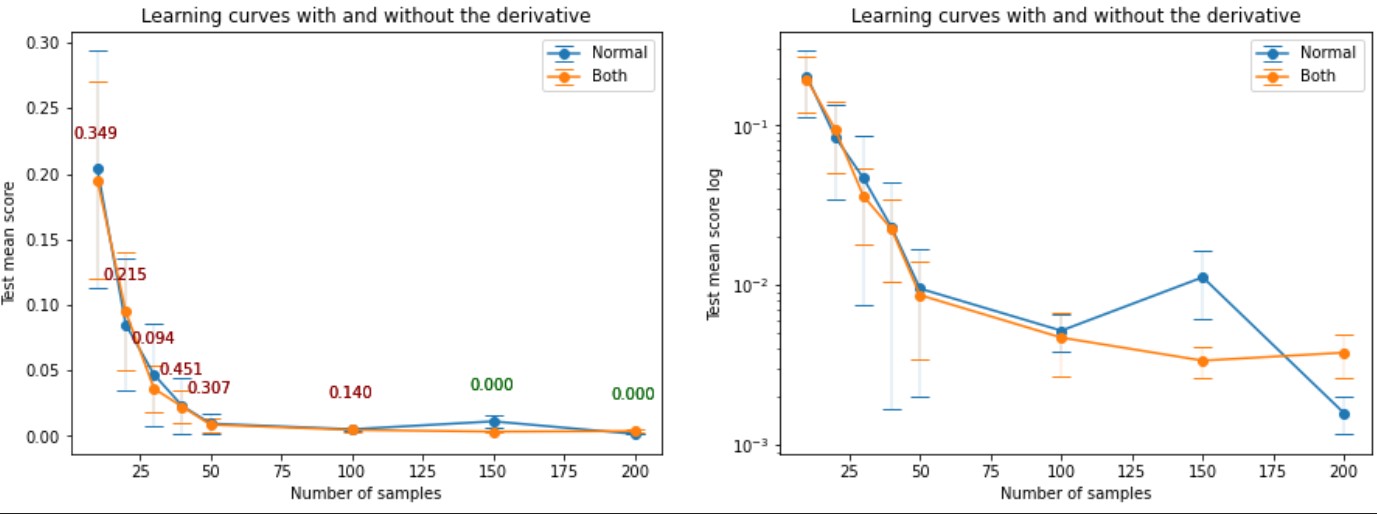}
    \centering
    \caption[Muti-task gradient learning on polynomial]
    {Muti-task gradient learning on polynomial - comparison between Multi-task gradient learning, or \textit{Both}, and \textit{Normal} approaches for Task 1 using different sample sizes for training. The graphic on the right is in log scale. Green and red annotation represent the p-value that, respectively, respect or not the threshold determining the equality of the models}

    \label{fig:FinalPolynomialBoth}
\end{figure}

In contrast, figure \ref{fig:FinalPolynomialBoth} shows as the multi-task learning approach was not advantageous.

\subsubsection{Task 2}

For both NACA datasets, in their turns, the data generation was limited. In fact, 65 simulations were used as a base for Task 2, in a way that 10 examples were randomly picked from them all and maintained fixed for testing. The rest was used to sample training examples, according to the need.




\subsubsection{Task 3}

Only 34 final adjoint simulations were produced, where 10 were fixed for testing and 4 for validation.

\begin{figure}[htb]
    \includegraphics[width=14cm]{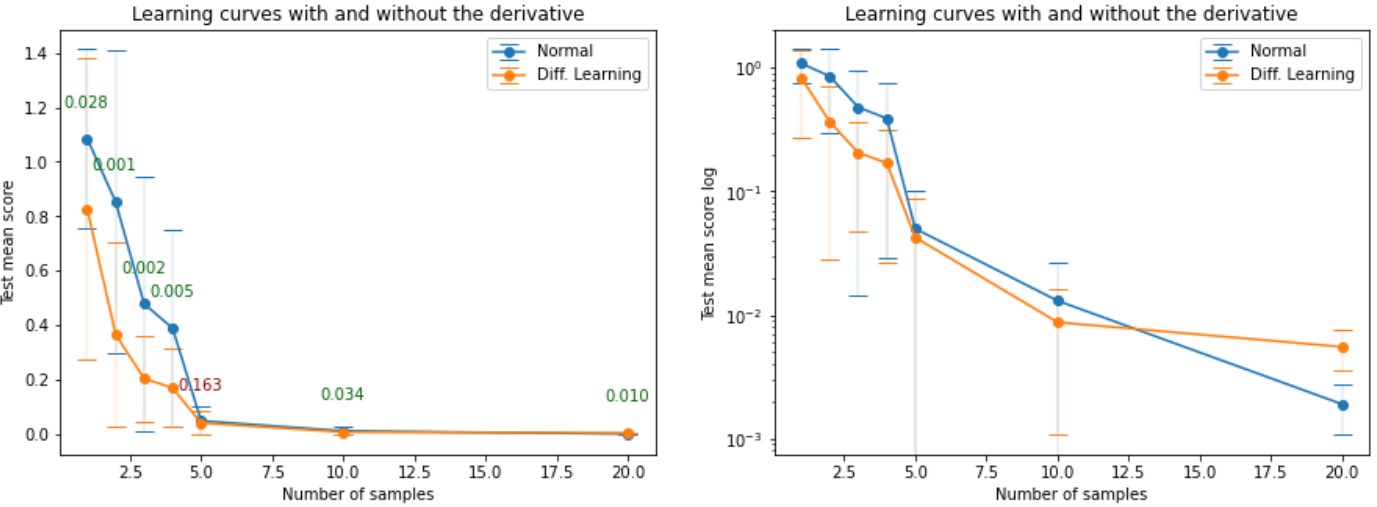}
    \centering
    \caption[Differential Machine Learning on the NACA dataset]
    {Differential Machine Learning on the NACA dataset - comparison between Differential Machine Learning and \textit{Normal} approaches for Task 3, using no gaussian filter. The use of gaussian filter resulted in a similar curve, but with better performance on small training sets and worse when more data was available. Green and red annotation represent the p-value that, respectively, respect or not the threshold determining the equality of the models.}

    \label{fig:FinalPolynomialNaca}
\end{figure}

Finally, figure \ref{fig:FinalPolynomialNaca} illustrates the power of Differential Machine Learning and how it is capable, with only two samples seen by the model, to achieve the same performance as using four examples. It also demonstrates how rich information is the adjoint in fluid dynamics. The graphics also indicate the p-value for the hypothesis $H_0$, meaning that when $p$ is under 0.05 it can be rejected with 95\% confidence and both errors are in fact different.

Moreover, as we can see, the less generalizable the original model is, the stronger will be the regularization effect. Nevertheless, when the model achieves a good performance, for example with 20 training samples, the approach stops working. This behavior is different from what was observed in the toy cases, where both curves came closer, and can be attributed to the fact that the gradient produced by the continuous adjoint is not exact.

\subsubsection{Task 4}

\begin{figure}[htb]
    \includegraphics[width=14cm]{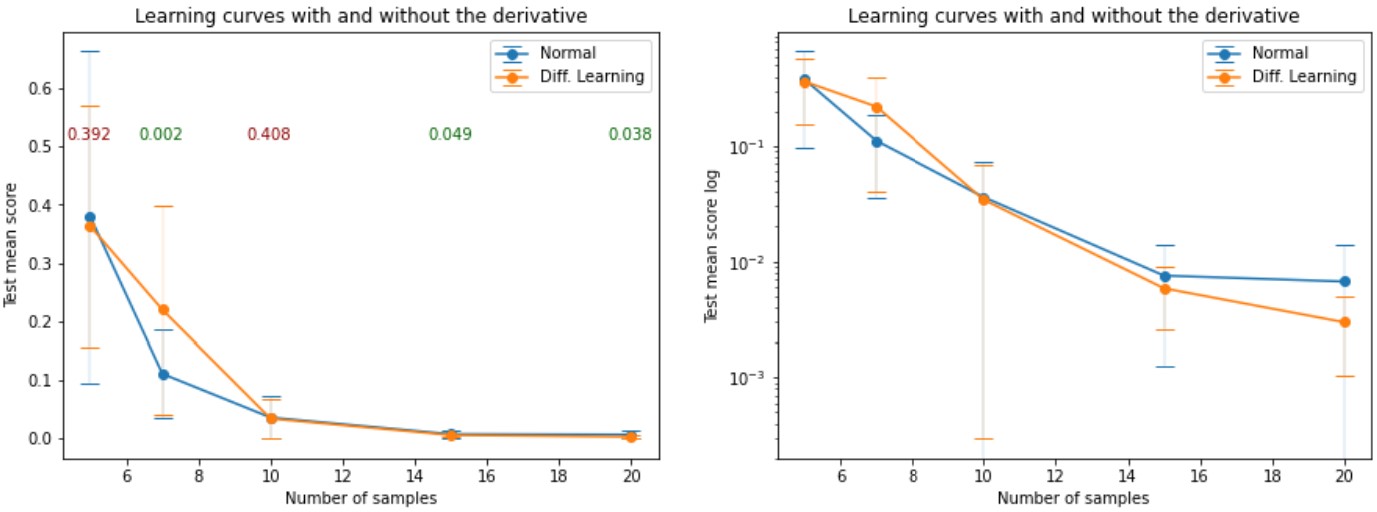}
    \centering
    \caption[Differential Machine Learning on the boundary conditions NACA dataset]
    {Differential Machine Learning on the boundary conditions NACA dataset - comparison between Differential Machine Learning and \textit{Normal} approaches for Task 4 using different sample sizes for training. The graphic on the right is in log scale. Green and red annotation represent the p-value that, respectively, respect or not the threshold determining the equality of the models.}

    \label{fig:FinalBC}
\end{figure}

The same number of simulations was produced and the same train-test-validation setup as in Task 3 was used. Results shown in figure \ref{fig:FinalBC} exemplifies how the framework can work even when no shape variation is seen by the model.

Even if the performance improvement doesn't achieve the same level as in Task 3, it is a surprising phenomenon that supervising with the geometrical gradient can help the model generalize to other boundary conditions. The intuition behind this characteristic is that, despite deriving w.r.t. the position, the objective function (lift) depends on the freestream velocity.

This can also open doors for geometrical extrapolation. Even if the available data doesn't contain much or no shape variation, including the adjoint gradient can add extra information.

\section{Assessing the model's performance on extrapolation task}

To verify the generalization improvements of Differential Machine Learning more clearly, another experiment can be done. The geometrical NACA dataset can be used, but instead of randomly sampling the evaluation examples, specific shapes were separated. More precisely, the 10 geometries with higher \gls{M} and \gls{P} digits, i.e., $M \in [4, 6]$. The training, in the other hand, only contained geometries within the range $M\in [0, 4]$, and \gls{P} working as a second sorting number. This is going to be called \textit{Task 5}.

\begin{figure}[htb]
    \includegraphics[width=12.5cm]{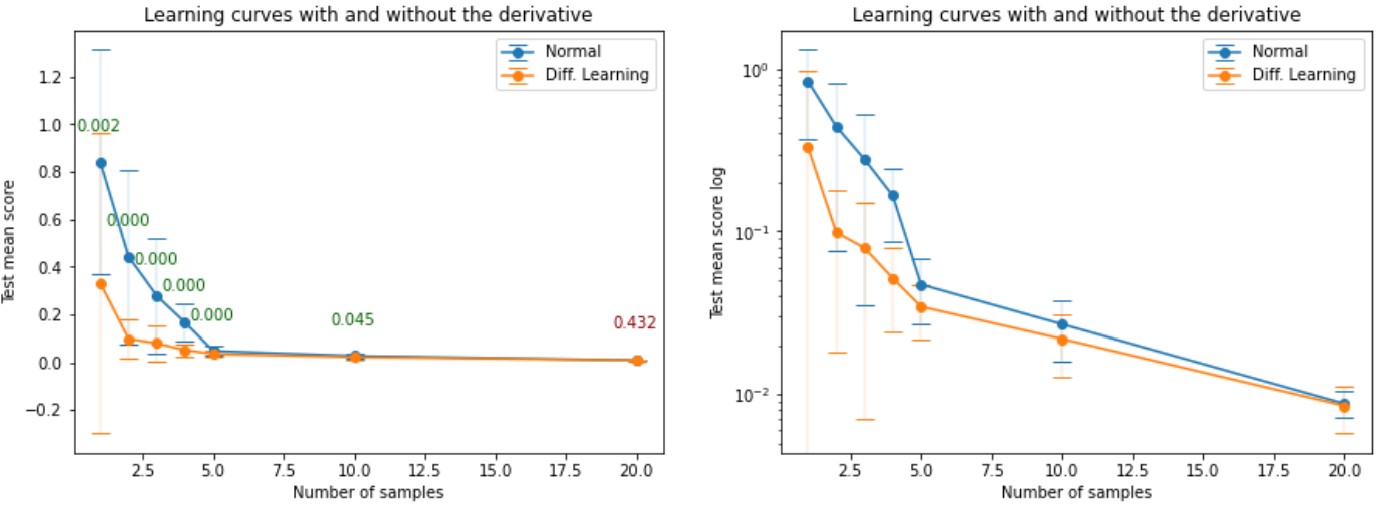}
    \centering
    \caption[Differential Machine Learning extrapolation on NACA dataset]
    {Differential Machine Learning extrapolation on NACA dataset - comparison between Differential Machine Learning and \textit{Normal} approaches for the NACA dataset, with different shapes in test and train - Task 5. Green and red annotation represent the p-value that, respectively, respect or not the threshold determining the equality of the models.}

    \label{fig:FinalPolynomialNacaGen}
\end{figure}

In this way, the network sees different shapes than it predicts, and it is further assessed the extrapolation power, instead of interpolation. The regularization effects (figure \ref{fig:FinalPolynomialNacaGen}) are even stronger.

\section{Ablation study}

 In this section, it is conducted an ablation experiment to analyze the effectiveness of different techniques used in the final framework for Task 3. The hyperparameters were optimized using the validation set. Each training of 10 samples was realized 20 times for different seeds, and the mean and standard deviation are summarized in table \ref{table:ablation}.
 
 We can observe that each technique, apart, is not responsible for a considerable effect. It is only when they are assembled that the framework achieves its potential.
 
 \begin{table}[H]
    \center
    \centering
    \begin{tabular}{c|cccc|c}
    
    \hline
    & Gradient & Cut & \gls{alpha} & Normal iters &
    Test error ($\times 10^2$) \\
    \hline
    1 & - & - & - & - & 1.34 $\pm$ 1.55\\
    2 & \checkmark & - & - & - & 1.95 $\pm$ 1.73 \\
    3 & \checkmark & \checkmark & - & - & 1.71 $\pm$ 1.75  \\
    4 & \checkmark & - & \checkmark & - & 1.08 $\pm$ 0.87\\
    5 & \checkmark & \checkmark & \checkmark & - & 1.04 $\pm$ 0.69 \\
    6 & \checkmark & \checkmark & \checkmark & \checkmark & \textbf{0.80} $\pm$ 0.71\\
    \hline

    \end{tabular}
    \caption[Ablation study on final framework with 20 runs]
    {Ablation study on final framework with 20 runs - where \textit{Gradient} refers to the Differential Learning framework, \textit{Cut} to remove the gradient from the trailing edge, \gls{alpha} to the scaling parameter (already optimized) and \textit{Normal iters} to 50 iterations without gradient before using Differential Learning. The combination of the columns was chosen to give a general idea of how each unit influences performance and how their combination is important. Each model was run 20 times with different seeds (sampling and initialization) and the mean and standard deviation are displayed on the table.}
    \label{table:ablation}
\end{table}

\section{Computational cost}

The computational cost of the approach presented in this work is heavier than the baseline. To begin with, an adjoint simulation takes in total, theoretically, twice the time of a primal simulation, because it has to resolve two \acrshort{PDE}s at each iteration.

In practice, however, the mesh had to be refined to gain precision for the adjoint, resulting in a slower simulation. In addition, the number of iterations done to the adjoint, \textbf{14300}, was much higher than the \textbf{5500} of the primal, to assure the gradient's quality. Each iteration, in its turn, presents multiple resolutions (a variable number of times) for each field variable, aiming to decrease the residues. Table \ref{table:timing} presents how much time takes, on average, for each iteration.

The Differential Machine Learning framework is also slower, as it backpropagates twice through the network.

\begin{table}[H]
    \center
    \centering
    \begin{tabular}{c||cc|cc}
    
    \hline
    & Primal & Adjoint & Normal PointNet & Diff. Learning PointNet \\
    \hline
    \textbf{Time (s/iters)} & 0.58 $\pm$ 0.4 & 2.13 $\pm$ 1.15 & 0.02 & 0.037 \\
    \hline

    \end{tabular}
    \caption[Timing for gradient regularization in fluid dynamics]
    {Timing for gradient regularization in fluid dynamics - on the left, time for each iteration of primal and adjoint, considering all the 34 final simulations; multiple resolutions are done per iteration according to the tolerance defined. On the right, time for a single epoch in one training with 3 examples on train, 5 on test and 3746 parameters; a stable value.}
    \label{table:timing}
\end{table}
  \chapter{Conclusion}

Using the gradient generated by the adjoint method has been proved to be an efficient method to boost the generalization performance of Deep Learning models for fluid dynamics. This work developed a reliable dataset for \acrshort{CFD} simulations with the gradient computed by the adjoint method. 

In addition, it proposed improvements to Differential Machine Learning and leveraged it to use in fluid dynamics and graph models, achieving important results, even when a high-quality gradient is not available. The products of this work allow, in some cases, the use of half the original quantity of data to achieve similar performances.

The use of gradient data in Machine Learning is still an area of research to be further explored. Specifically its use for fluid dynamics modeling shows a still unexplored potential. In this context, some interesting directions of research arise, following this work. 

\textbf{Future works} would include the generation of a new dataset using the discrete adjoint approach, as it generates an exact gradient. This would possibly function even better to apply Differential Machine Learning. More reliable results could also be achieved by implementing a Bayesian hyperparameters optimization, instead of a grid search. Finally, different, more complex deep learning architectures could be tried, rather than PointNet. GAT \cite{velickovic_graph_2018} could be a good alternative for a graph model.

\textbf{A promising direction} of research, based on the results achieved, would be the task of shape optimization. As it was proved along the text, the gradient of the \textit{DiffLearning} model respects the differential labels, which would possibly yield more correct sensitivity maps.

Despite the good outcome, the research was \textbf{limited} by the small datasets used as a consequence of the high computational cost of adjoint simulations. Correlated, the precision of the gradient highly depended on resources available due to the continuous adjoint strategy.

  \backmatter
  \nocite{*}
  \bibliographystyle{poli-unsrt}
  \bibliography{thesis}
  \appendix
  \chapter{Convolution Neural Network models for fluid dynamics}
\label{appenA}
\gls{CNN} are a type of Neural Networks that implement a convolution layer, helping to change the dimension of the feature space in a 3d fashion. Because of this characteristic, it is widely used for image tasks, where the input data has a 3d representation, acting as filters to automatically extract features.

Luckily, a 2d or 3d mesh representation can benefit from CNNs in a similar way, especially a Cartesian grid. In this way, \cite{thuerey_deep_2020} applies \acrshort{CNN}s to model 2d \acrshort{CFD} simulations (using Euler equations \ref{EulerEquations}) of the airflow around an airfoil shape. This appendix is based on this article. Differently from other Machine Learning strategies for this modeling, which target a simplified representation of the input as parameters of interest \cite{yu_flowfield_2019}, the work in question infers it directly, i.e., the actual quantities in each cell of the domain.

The data is generated by changing the boundary conditions dictated by the Reynolds number, and varying the \gls{aoa} of the airfoil. Then, the solution of the \acrshort{CFD} simulation is computed and stored. The range and nature of these inputs depend on what are the goals of the \acrshort{DL} model, because it will generalize better to data similar to the one it has been trained on. Another possibility, for example, is to include data from different airfoil shapes, so the model can generalize better to new geometries. It is important to understand, nevertheless, that an enlarged space of solutions yields also a more difficult learning task, and thus worse performance for the model.

\subsubsection{Data used in the Network}

Once the simulations' results are available, the input and output data have to be retrieved for the Network training and testing. To achieve this, some pre-processing steps are necessary to transform the simulation grid into the grid used by the architecture.

\begin{figure}[htb]
\includegraphics[width=15cm]{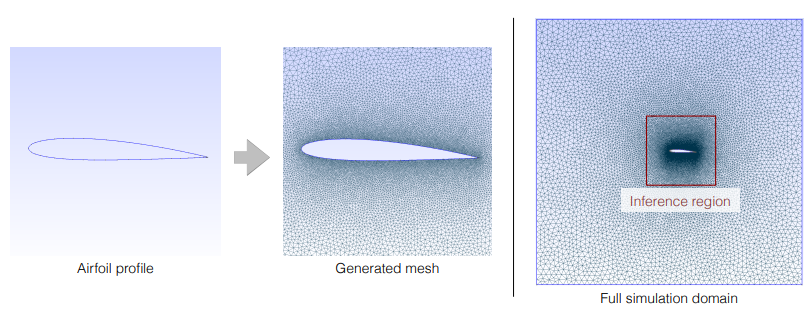}
\centering
\caption[ Inference Region ]
{Inference Region - procedure to mesh the domain and reduce it to the region to be modeled by the network \cite{thuerey_deep_2020}.}
\label{InferenceRegion}
\end{figure}

This is fundamental for two reasons. First, the domain simulated is usually much bigger than the actual region we are interested in (around the geometry), and therefore this domain should be cropped to avoid useless information (figure \ref{InferenceRegion}). Second, the original mesh changes each time the geometry changes (different or shifted shape) in order to adapt to it, and thus it is necessary to interpolate the results to a fixed grid, as the Network has to receive always the same input size. In this way, the shape of the input Cartesian grid is 128x128.

Then, the actual input to the \acrshort{CNN} can be chosen. To account for the geometrical shape and the initial flow conditions, the input has the shape 128x128x3. It encodes in the first two channels the velocities in x and y initialized to the constant correspondent free stream condition and zero value inside the airfoil shape, while the third channel is a binary mask of the geometry. This setup is intentionally highly redundant, including three times the geometry and repeating the constant free stream value to improve the capability of the model to learn this important information - known to directly affect the result.

The output of the Network has the same shape, which means it is predicting values for the whole grid as previously stated. The channels, in its turn, are the pressure and the velocity in x and y from the \acrshort{RANS} solution.

\subsubsection{Important pre-processing steps}

Besides the treatment already implemented in the database, some more modifications are important to ensure better performance of the \acrshort{CNN} and the representation of the physics. Common in Deep Learning in general, is the normalization of all quantities, i.e., the input data and the \textit{true} target outputs, in order to flatten the space of solutions.

This can be done w.r.t. the magnitude of the free stream velocity, in a way that $\widetilde{v}_{0}=v_0 / |v_i|$ and $\widetilde{p}_0=p_0 / |v_i|^2$. The quadratic term in the latter is important to remove the scaling of the same order of the pressure from the ground truth label. This transformation also makes the variables dimensionless, simplifying the space of solutions.

Another correction can be understood by analyzing equations \ref{NvSkSimplified}: only the gradient of the pressure is necessary to compute the solution, and the same usually happens with \acrshort{RANS} equations. Therefore, by removing the mean pressure from each value ($\hat{p} = \widetilde{p}_0 - p_{mean}$) we assure this quantity is correlated to the input and random pressure offsets will not affect this relationship.

The above pre-processing techniques were tested and proved to help to achieve better results. Finally, all channels are again normalized, this time to scale to the range [-1,1] aiming to reduce errors from limited numerical precision.

\subsubsection{Convolution Neural Network architecture}
\begin{figure}[htb]
\includegraphics[width=15cm]{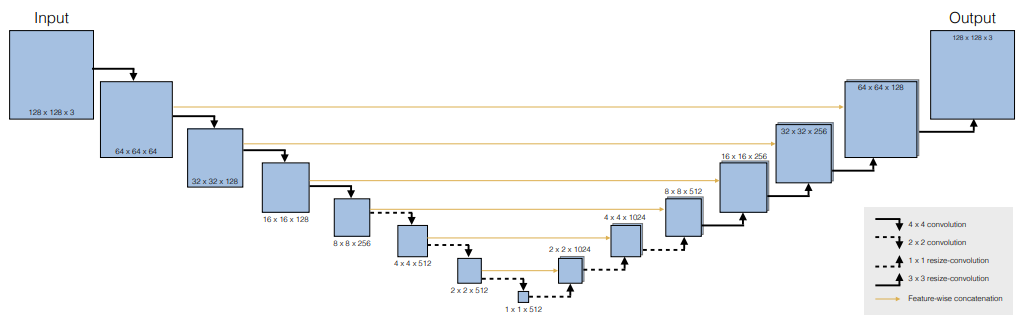}
\centering
\caption[ U-net architecture ]
{U-net architecture - DL architecture where the first set of layers is responsible to reduce the original feature space to a latent representation. This procedure allows regrouping local patterns of information. The second half mirrors the behavior, retrieving the initial shape
\cite{thuerey_deep_2020}}
\label{Unet}
\end{figure}

Many complex architectures would be possible for this application by varying the number of layers, kernel sizes, strides and pooling layers used. A simple architecture, already validated by previous works in terms of error and memory performance, was chosen as a base: the U-net. Figure \ref{Unet} shows how this architecture works by down-sampling the grid size and increasing the feature channels (encoding part) with convolutional layers until it gets to a small image size. Then, the behavior is mirrored with average-depooling layers and returns to the final shape, assuring it is the same as the initial, as needed.

Batch normalization and non-linear activation layers are also used at each step. Moreover, long skip connections are used in the U-net to recover fine-grained information from the encoding part to the decoding. The model can be achieved by trying different skip-connections and by changing other hyperparameters such as the batch size, the learning rate and learning rate decay of the optimizer (Adam), and the number of weights of the architecture (scaling the feature maps of the convolutional layers).  

Regarding the classical techniques to avoid overfitting, only a small amount of dropout was chosen because it was noticed to improve the results. Other methods like data augmentation or weight decay, did not apply to the problem or were less important to \acrshort{CNN}s.

As metrics, the L1 loss ($|y_{predict} - y_{true}|$) was used for presenting slightly better results in tests than the L2 loss. On the other hand, the evaluation between models is the average relative error for all inferred fields ($e_{avg} = (ep + e_{vo,x} + e_{vo,y})/3$), taking into account all fields. Ideally, cross-validation strategy should be used to train/test the model.

\subsubsection{Conclusions}

\begin{wrapfigure}{l}{8cm}
\includegraphics[width=8cm]{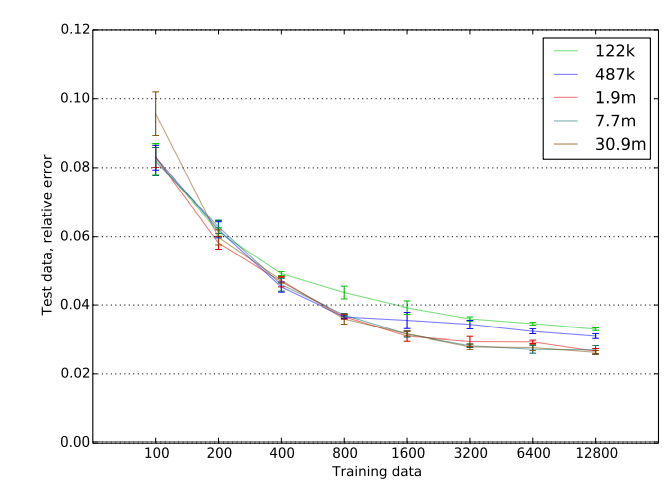}
\centering
\caption[ Learning curve for models of different sizes]
{Learning curve for models of different sizes - different model sizes and training data amounts: more complex models overfit with fewer samples and saturates on a lower error when more samples are available
\cite{thuerey_deep_2020}.}
\label{AccuracyCNN}
\end{wrapfigure}

The first important conclusion the article does is regarding the amount of training data used. As it was to be expected, the more data used, the smaller is the error of the model, stabilizing when getting to a very large amount as the model saturates and lacks complexity to describe all particularities existent.

Related to the first point, is the number of weights of the network, which is directly tied to the complexity of the network. A high complexity network can better reproduce the complexity of the dataset when enough observations are available, in a way it saturates later than simpler models. In the other hand, it can underfit the data when a smaller number of observations can be used, being preferable to use the networks with fewer weights. Figure \ref{AccuracyCNN} expresses well this behavior.

This represents a clear trade-off between the performance of 
the network and the computational cost. As it was stated in the introduction section of this work (chapter \ref{Introduction}), \acrshort{CFD} 
simulations are quite expensive to solve, but the more this is done, the better the process can be represented by Deep Learning. 

Notwithstanding, the best model found had only a 3\% relative error, while others were less accurate, but also yielded reasonable results, proving that modeling \acrshort{CFD} simulations with \acrshort{CNN}s can be a very interesting approach.

  \chapter{Geometric DL: graphs and manifolds}
\label{appenB}
Unstructured data cannot be directly used in \acrshort{CNN}s, which demand a structured representation, like images, for example. Therefore, when analyzing \acrshort{CFD} surface variables originated from irregular meshes, they cannot be used. This is one kind of geometric data, which is similar to a \textit{Point Cloud}. Point Cloud is a type of geometric data representation possessing an irregular format, i.e., point coordinates in space written in an unordered manner to represent a 2D/3D shape design. 

Thus, the irregular mesh can be considered equivalent to it if only the cell centers, or cell nodes, are taken into consideration. The following articles deal with \textit{Point Clouds} and could be applied as well in regression tasks on meshes. The data could also be converted to a structured grid (voxelization - appendix \ref{appenA}), but this can make this information unnecessarily voluminous and carry approximation issues from the transformation.

\section{PointNet}

PointNet \cite{qi_pointnet_2017}{} is an architecture built to work with Point Clouds. It takes Point Clouds in euclidian space as input (plus point features - global or individual) and is able to predict a single value (image classification or regression of a performance variable) or a value per point (image segmentation or regression of surface variables).

As a first principle to respect, PointNet should take into consideration the fact that Point Clouds are invariant to permutation of the points, as well as rotation and translation of the whole dataset. The last two characteristics are less important in our case since the data is stemming from CAD models and have an invariant representation. Figure shows the scheme for PointNet.

\begin{figure}[htb]
\includegraphics[width=\textwidth]{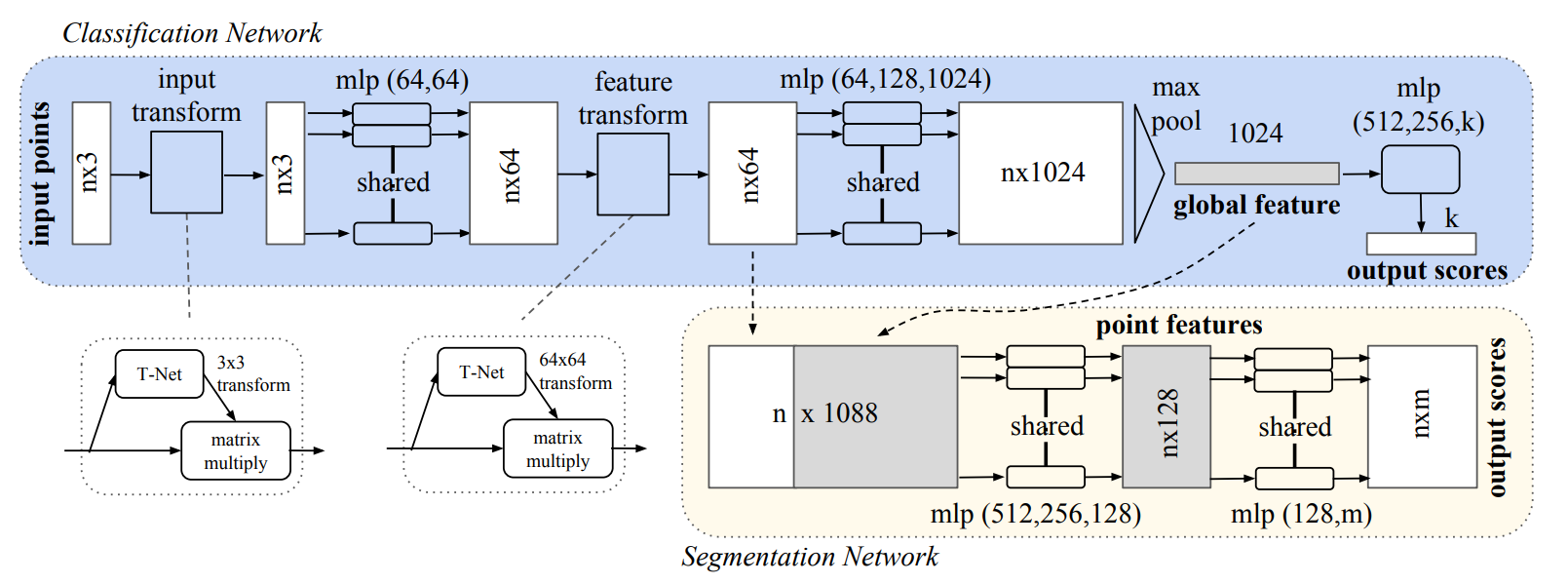}
\centering
\caption[PointNet architecture]
{PointNet architecture - units for both classification or segmentation
\cite{qi_pointnet_2017}.}
\label{PointNet}
\end{figure}

The max pooling operation shown on the scheme is fundamental to extract the global features of the data and, at the same time, to make the model invariant to input permutation, because of the symmetry of this function. 

It is important to notice that the segmentation and classification tasks require different procedures, due to their different outputs. For the latter, the global features can be used directly in an \acrshort{MLP} to predict the labels (or performance variables). For the segmentation and surface variables prediction, however, the output should be per point, and thus the global features are encoded in the original point vectors. As previously stated, however, models for global performance variables should be avoided, being preferable to integrate the over the surface.

The T-Net mini-networks at the beginning of the architecture serve to make input invariant to transformations (rotation and translation).

\section{PointNet++}
PointNet++ is an improved version of PointNet. It has the exact same goals, but corrects the problem of not capturing local structures of the data - the common characteristics of points in the same spatial region. This can limit generalization performance and the recognition of patterns; different from \acrshort{CNN}s, that can identify these local features.

To achieve this goal, PointNet++ is a hierarchical neural network, meaning it partitions the original set of points into overlapping clusters, and repeats this process more than once. These clusters are neighborhood balls in the Euclidian space, represented by its scale and centroid. They are generated by the farthest point sampling (FPS) algorithm.

The centroids of the local regions are generated by \textit{Sampling Layers}, which use FPS to sample points from the input. Then, the \textit{Grouping Layer} finds the local regions around the centroids within a radius, using a Ball Query algorithm, and encoding its \textit{K} points, which vary according to the group. In addition, to improve generalization performance for cases where the density of the Point Cloud is irregular, containing sparse regions, PointNet++ proposes extracting the groups with multiple scales, according to local point density. Next, a regular PointNet (semantic type) is used to extract and encode the patterns as we previously saw, and can properly do it despite the inconstant number of points \textit{K} per region. The coordinates of the region points are transformed previously to be relative to the centroid.

\begin{figure}[htb]
\includegraphics[width=\textwidth]{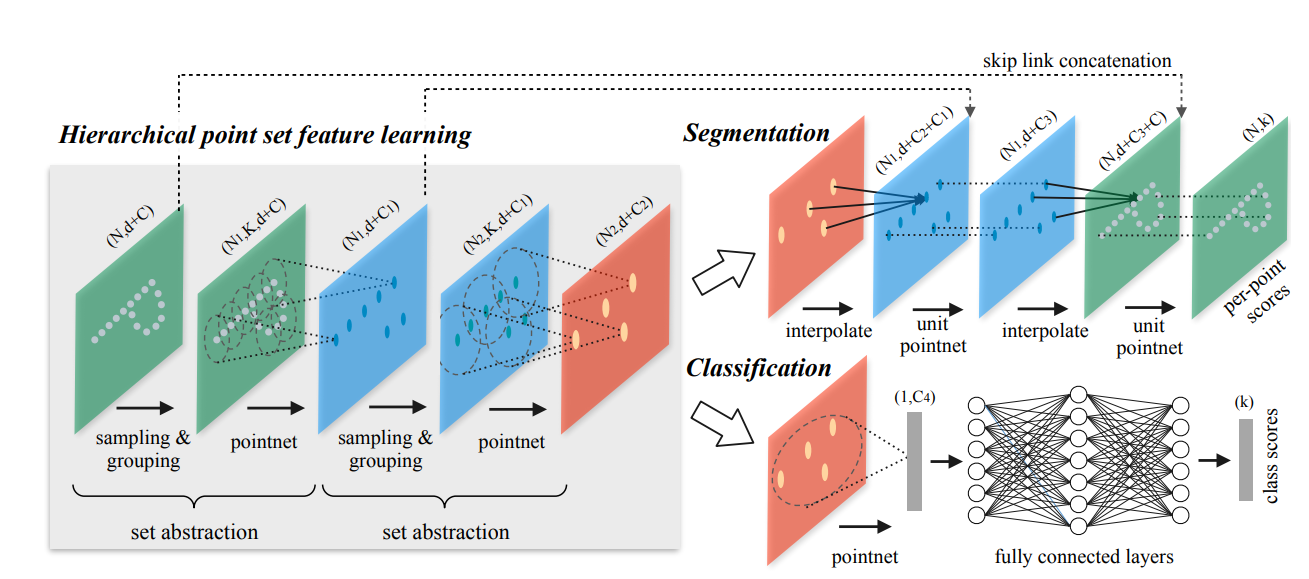}
\centering
\caption[PointNet++ architecture]
{ PointNet++ architecture - units for both segmentation and classification, including PointNet layers
\cite{qi_pointnet_2017-1}.}
\label{PointNet++}
\end{figure}

As in PointNet, the network task can be of classification or per point segmentation, which are respectively equivalent in regression to computing performance variables and surface variables. In the first case, the architecture remains simple after the feature learning and uses a single PointNet. For the second one, however, as the points are in a reduced representation after the feature learning, an interpolation is used to pass to the next hierarchical level. Then, the features are concatenated with the corresponding representation of the feature learning stage. A unit PointNet finalizes the procedure to get to the next hierarchical level.

\section{Graph Attention Network}
The \gls{GAT} \cite{velickovic_graph_2018} is an architecture that has a different approach from the previous ones. It uses the concept of graphs, with nodes and edges, to represent the system. In comparison with other Graph Networks, however, it does not need any previous knowledge about the original graph structure, as it computes the edges itself.

The architecture is based on special Graph Attentional Layers, which are stacked to form the whole network. This type of layer performs a series of operations in the nodes, which, in our case, are the points of the Point Cloud.

The input is the same as before, the set of points with their coordinates and individual/global features (dimension \textit{F}). To increase dimensionality a linear transformation is applied, using the same matrix W for each multiplication to achieve dimensionality \textit{F}’. The edges are then calculated, only between nodes of the same neighborhood. This neighborhood can be determined by previous knowledge of the graphs, for example which cells are connected to each other. They are scalar \textit{attention coefficients} \cite{wang_attention_2021}.

\begin{equation*}
    e_{ij} = a (\mathbf{W}\Vec{h_i}, \mathbf{W}\Vec{h_j})
\end{equation*}

Where \textit{a} represents a shared parameters LeakyRelu Layer applied to the matrix formed by the concatenated new vectors. The edge values are then normalized with a Softmax function, to make them comparable with each other: $\alpha = Softmax_j(e_{ij})$. Finally, a linear combination of the normalized edges (attention coefficients) with its correspondent features is done, applying also some non-linearity to achieve the final output of the node.

\begin{equation*}
    \Vec{h}_i' = \sigma (\sum_{j \in N_i} \alpha_{ij}\mathbf{W}\Vec{h_j})
\end{equation*}

Moreover, these features per node can be concatenated with the ones from other nodes, 
increasing the dimensionality, except in the final layer. In this case, averaging can be done:
 
\begin{equation*}
    \Vec{h}_i' = \sigma (\frac{1}{K} \sum_{k=1}^K \sum_{j \in N_i} \alpha^k_{ij}\mathbf{W^k}\Vec{h_j})
\end{equation*}

As the outputs are always per node/cell, we should integrate it over the surface to find the global variables or use a PointNet at the end.

Many other \acrshort{GNN}s exist and are suitable to work with geometric data. Simpler and efficient options are GraphSage \cite{hamilton_inductive_2018} and Graph U-nets \cite{gao_graph_2019}, while other more complicated can even use \textit{transformers} \cite{pan_3d_2021}.
  \chapter{Graph Models for fluid dynamics}
\label{appenC}

\glspl{GNN}, are another type of Deep Learning architecture that can be successfully employed to model \acrshort{CFD} simulations. They map the input as a graph structure, which contains nodes with connections between them, and compute dynamics via learned message-passing to output another graph that can be converted to the targeted quantities. \acrshort{GNN}s can be compared to \acrshort{CNN}s in a way that, despite each element being updated as a function of its neighbors, this neighborhood is not fixed anymore, but varies and have connected edges.

\begin{figure}[htb]
\includegraphics[width=8cm]{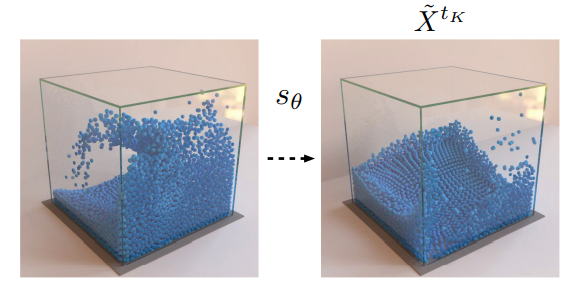}
\centering
\caption[\acrshort{GNS} simulations]
{\acrshort{GNS} simulations - particle simulation and update from one time step to another \cite{sanchez-gonzalez_learning_2020}.}
\label{ParticleSimulation}
\end{figure}

Article \cite{sanchez-gonzalez_learning_2020} implements a model called \gls{GNS}, that generates a robust and generalizable result. It targets, however, dynamic mesh-free particle simulations (figure \ref{ParticleSimulation}), where the physical system is described as particles that interact with each other, different from the meshed simulations we target in this work, for example using a Finite Volume Analysis. Notwithstanding, \acrshort{GNS} can also be applicable to those simulations, expressing, instead of the particles, the cells of the mesh as the nodes of the graph.

\subsubsection{Graph Neural Simulator model}

As the \acrshort{GNS} model focuses on particle-based simulations, it represents states ($X^t$) as a group of particles, each one encoding mass, material, movement and other properties at a certain time step. The dynamics are computed to achieve the subsequent time step, and the set of all these states describes the trajectory of the particles. Again, this approach intends to model the small-time particularities of the flow, being different from the \acrshort{RANS} modeling that was exposed until now in this document.

The machine learning framework is used to make these updates between timesteps, in a way that $X^{t_{k+1}} = s(X^{t_k})$. In each of these timesteps, each particle has encoded its individual state $x_i$, and joining them forms the general state of the domain. The learned interactions between the particles/nodes/cells are what will determine if the modeling of the CFD simulation yields a quality and generalizable solution. The simulation can be seen as the message-passing of the graph network, where nodes are particles or cells, and edges correspond to pairwise relations between them.

The individual state $x_i$ of the particle stocks the following information: its position, its last 5 previous velocities (number found by hyperparameter search) and static properties of the material. It should be noticed, once again, that this state is based on particle-based simulations and should be different according to the kind of simulation used; linking with \cite{thuerey_deep_2020}, it could be used the corrected velocities (in x and y) and the pressure at each cell.

\begin{figure}[htb]
\includegraphics[width=15cm]{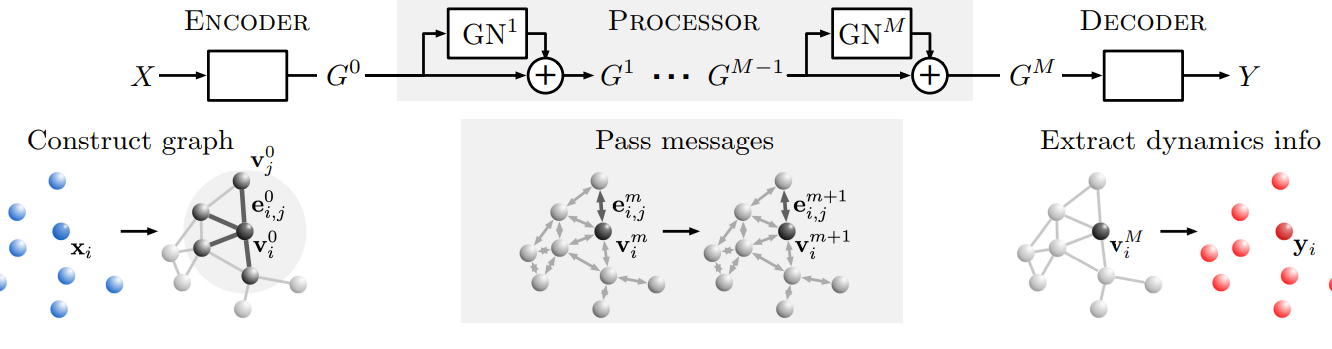}
\centering
\caption[ \acrshort{GNS} scheme ]
{\acrshort{GNS} scheme - architecture to model a complete rollout of particle simulations, GN being a Graph Netwotk layer, 
\cite{sanchez-gonzalez_learning_2020}.}
\label{GraphNetwork}
\end{figure}

To implement the graph learnable system, there are 3 necessary steps: Encoder, Processor and Decoder (represented in figure \ref{GraphNetwork}).

\textbf{Encoder}

The Encoder is responsible to construct the graph scheme. It takes as input the state $X^t$, thus all the information in the current step of the simulation, and converts it to nodes ($v_i \in V$) and edges ($e_{ij} \in E$). This embedded, latent graph will be called $G_0$. The node embedding can be represented as $v_i = \epsilon^\nu (xi)$, and thus are learned from particles' states. The edges, in the other hand, are the representation of the interaction between the nodes, represented as $e_{i,j} = \epsilon^e (r_{i,j} )$. In other words, they can be learned from the pairwise properties, such as displacement, for example, from particles $i$ and $j$: $r_{i,j}$. 

The learnable functions $\epsilon^e$ and $\epsilon^\nu$ are implemented as \gls{MLP}. One last embedding is done to represent $u$ the global properties of the system, if applicable, like gravity, and that can also be directly included as properties of a node.

Before embedding the edges, however, one piece of information is necessary: to know which nodes will interact with each other. This can be done by the means of a \textit{Nearest Neighbors} algorithm using a connectivity radius \textit{r}. This value is considered a hyperparameter of the model.

\hfill \break
\textbf{Processor}

The Processor is responsible for the actual modeling of the simulation and its physical system. In practice, it takes as input the initial graph $G_0$ and updates it in $M$ learned message-passing steps, computing, thus, the interaction between nodes. Each update can be represented as $G^{m+1} = GN^{m+1}(G^m)$, until it gets to the final graph $G^M$. The number of message-passing updates is a hyperparameter of the model and should increase to represent the complexity of the system, correctly propagating the necessary physical information.

Therefore, the Processor can be considered as a stack of Graph Networks. Each GN is, like for the encoder, an \acrshort{MLP} with identical structure and unshared parameters, causing the final model to have \textit{M} times more parameters as it resembles a deep architecture. This option yielded a better accuracy result in comparison with a shared parameters stack of \acrshort{GNN}s, which have a strong inductive bias but did not have a significant advantage concerning overfitting or computational cost.

\hfill \break
\textbf{Decoder}

Finally, the decoder mirrors the behavior of the encoder to get to the final result. In general lines, it will take as input the final graph $G_M$ and transform it into dynamical information $Y^t$, comparable with a state $X$. Once more, an \acrshort{MLP} is used for the decoding function.

The output used in the article, for example, was the acceleration per-particle. This output acts also as the target value for the supervised learning training, as it is compared to the ground truth solution (calculated by finite differences from the position). Having the acceleration of each particle, it is possible to use semi-implicit Euler integration to compute the position and velocity used in $X^{t+1}$. Other outputs could also be contemplated according to the type of simulation, for example the velocities and pressure per cell described in \cite{thuerey_deep_2020}. 

In the particle-based, time-dependant representation the update of the dynamical information to the next state $X^{t+1}$ ends the iteration, permitting the continuation of the loop to compute all the trajectory.

\hfill \break
\textbf{MLP details}

The \acrshort{MLP}s are a simple Neural Network architecture with only two hidden layers with ReLu activation. They all use also the same size and layer normalization (except the decoder), which addresses the drawbacks of batch normalization and is more suitable in this case, improving training stability.

\subsubsection{Pre-processing and training}

As it was done for \acrshort{CNN}s, normalization is also important in this case. All input and output of the \acrshort{MLP} is normalized online during each time step, resulting in faster training.

Another pre-processing technique was to include noise in the input. When modeling long trajectories, as models are trained step by step, error can be accumulated at each time step, generating a bad result at the end. This is because the error from the last time step confuses the model in the next one since it was never presented with such noise. The solution is to insert this noise in the input velocities, correcting position and acceleration when necessary. The amount of noise used is an important hyperparameter. In \cite{han_predicting_2022}, this strategy was replaced by a transformer-based architecture.

For training, the strategy was similar to the \acrshort{CNN}s - choosing the L2 loss, using Adam optimizer and varying its hyperparameters. A big difference is that, since the model is supposed to learn the dynamics of the particles, it trains in a single simulation, using mini-batches of particles. For evaluation and the choice of hyperparemeters, \textit{MSE} was used per time step in a training and test dataset.

\subsubsection{Conclusions}
In the evaluation analyses, \acrshort{GNS} presented a great generalization performance. Surprisingly, hyperparameters from the \acrshort{MLP}s themselves, such as the number of layers and its size, have little impact on the result. Instead, the number of message-passing steps makes a lot of difference, adding more complexity to the model. The connectivity radius (the larger the better) was important for the same reason, and the corruption of the input data with noise was also very relevant.

  \chapter{Deep learning in Computational Fluid Dynamics analysis}
\label{appenD}
Both \cite{sanchez-gonzalez_learning_2020} and \cite{thuerey_deep_2020} used an evaluation metric based on the cell/particle's exact position, comparing the output from the network with the ground truth solution. They both stress, however, that this approach can be limited when accessing the quality of the \acrshort{CFD} simulations modeling, because small shifts of the result, despite not reflecting bad physics, issue higher errors.
 
\begin{figure}[htb]
\includegraphics[width=13cm]{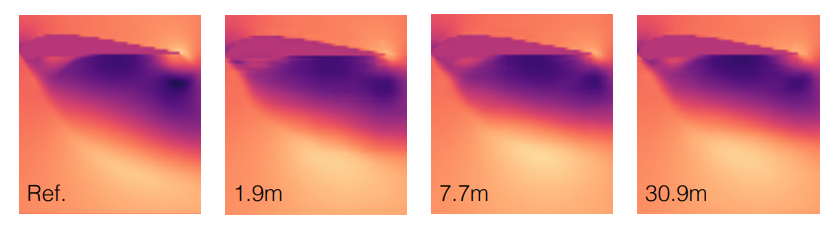}
\centering
\caption[Deep Learning sharpness in fluid dynamics]
{Deep Learning sharpness in fluid dynamics - three solutions for different network sizes. Despite having almost identical test error performance, solutions of the larger networks (on the right) are sharper \cite{thuerey_deep_2020}.}
\label{sharpness}
\end{figure}

Moreover, vector norms as loss functions can also cause bad conclusions. This is because small-scale changes in the result become barely noticeable, even if they derive a sharper solution, as shown in figure \ref{sharpness}. Therefore, looking at the result itself and comparing it with the ground truth solution is one important validation step.

  \chapter{Implicit and continuous neural representations}
\label{appenE}

To better model higher frequency functions, which is the case of the surface fields we are interested in, different alternatives are present. The following articles explore how to do it in the context of video reconstruction, but this can easily be used in other cases, much like this work.

\section{Neural radiance fields}

Neural Radiance Fields, or NeRF \cite{mildenhall_nerf_2020}, aims to apply \acrshort{MLP}s to directly map a point in space (5d scene representation) to its RGB value, thus being able to build a 3D scene from pixel images. The main idea could also be extended to a Point Cloud.

Modeling the original 5d coordinates representation with the \acrshort{MLP} yields poor results rendering high-frequency variations in color (output) or geometry (input). The solution proposed is called positional encoding, and it is based on the idea that “mapping the inputs to a higher dimensional space using high frequency functions before passing them to the network enables better fitting of data that contains high frequency variation”.

Positional encoding will apply the following transformation to each feature:
\begin{equation*}
    \gamma(p) = (\sin(2^0 \pi p), \cos(2^0 \pi p), \cdots, \sin(2^{L-1} \pi p), \cos(2^{L-1} \pi p)
\end{equation*}

Where p and L are hyperparameters to be tuned (L can even be different between features).

\section{Fourier features}

The same concept is expanded in \cite{tancik_fourier_2020}. Instead of a simple positional encoding, the feature transformation takes a more general view.

\begin{equation*}
        \gamma(v) = (a_1\cos(2 \pi b_1^Tv), a_1\sin(2 \pi b_1^Tv), \cdots, a_m\cos(2 \pi b_m^Tv), a_m\sin(2 \pi b_m^Tv))
\end{equation*}

It shows that doing this Fourier feature mapping for \acrshort{MLP}s enables learning high-frequency functions in low-dimensional problem domains. Notoriously, these \acrshort{MLP}s take low-dimensional coordinates as input and output a property at each location. By tunning the frequency parameter $b$, one can modify the range of frequencies that can be learned by the model.

In practice, better results are achieved by using $a_j$ =1 and $b_j$ sampled from a random distribution. Thus, the hyperparameter to tune would be the standard deviation of this distribution, normally Gaussian (despite the function being much less important than $\sigma$). A wider distribution results in faster convergence for the high-frequency components, yielding better results (in imagery tasks, it results in higher definition). On the other hand, a very wide distribution can cause artifacts in the result (noisy image). It is an underfitting/overfitting situation.

\section{Sinusoidal representation networks}

\acrshort{SIREN}s \cite{sitzmann_implicit_2020}, or Sinusoidal Representation Networks, use periodic activation functions, such as the sinus, in order to achieve similar results as the last articles. It is applied to the same point-based \acrshort{MLP}s, which have the problem of modeling the details of the underlying signals, especially if they are high-frequency.

\begin{equation*}
    x_i \longmapsto \phi_i(x_i) = \sin{W_ix_i + b_i}
\end{equation*}

Beyond this characteristic, \acrshort{SIREN}s are also differentiable, having continuous gradient functions that can be computed analytically. That happens because the derivative of a \acrshort{SIREN} is itself a \acrshort{SIREN}, as the derivative of the sine is a cosine: a phase-shifted sine. The same quality is not preserved by other common activation functions; ReLU, for example, has a discontinuous derivative and a zero everywhere second derivative. Some other functions do present this desired capability, such as Softplus, Tanh or ELU; however, its derivatives can be not well-behaved and fail to represent the fine details searched.

Therefore, they are well-suited to represent inverse problems, like the \acrshort{PDE}s we are so interested in. They can supervise the derivative of complicated signals, resulting in well-behaved predictive functions. Furthermore, \acrshort{SIREN}s were proved to converge faster than other architectures.

To achieve the desired results with \acrshort{SIREN}, an initialization scheme is necessary. It is used to preserve the distribution of activations through the network so that the final output at initialization does not depend on the number of layers. The solution is to use a uniform initialization of form 
$w_i \sim U(\sqrt{6/n}, \sqrt{6/n})$ so that the input to each unit is normally distributed with a standard deviation of 1. Moreover, the first layer of the network should span multiple periods over $[-1, 1]$, which can be achieved using $w_0=30$ in $\sin(\omega_0 \cdot Wx + b)$; this value should be changed depending on the modeled function frequency and the number of observations.

\section{Modulated sinusoidal representation networks}
The main issue with \acrshort{SIREN}, and also Fourier Features, is its generalizability. The network is optimized for a single signal of a specific frequency, and when the signal changes, having a different frequency, the model fails to represent 
it well. In practice, this architecture overfits the model to each new instance; thus, using multiple signals only performs well in low-resolution. The goal of modulation \cite{mehta_modulated_2021} is to make a single model generalizable to different instances in high-resolution, using only a single forward pass.

The network consists of two \acrshort{MLP}s connected to the functional mapping: the modulator and the \acrshort{SIREN}. The signal is transformed, before being used, in a low-dimensional (256) latent code $z$, acting as conditioning variables.

\begin{figure}[htb]
\includegraphics[width=13cm]{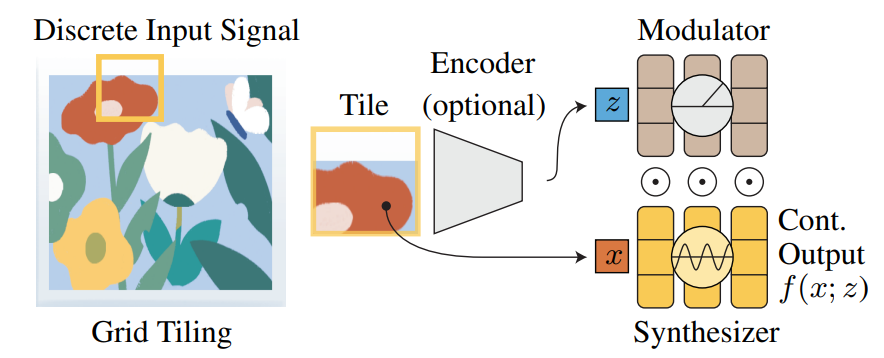}
\centering
\caption[Modulated \acrshort{SIREN} architecture]
{Modulated \acrshort{SIREN} architecture - tiling the input signal and link it to a SIREN network where each layer suffer modulation
\cite{mehta_modulated_2021}.}
\label{ModulatedSiren}
\end{figure}

As we can see from image \ref{ModulatedSiren}, the modulator uses the latent code directly, in order to output, for each layer, a parameter $\alpha$ that changes the amplitude, phase and frequency of the activations used in the \acrshort{SIREN} network. Mathematically, we have:

\begin{equation*}
    h_i = \alpha_i \circledcirc \sin{w_i h_{i-1} + b_i}
\end{equation*}

Where $\alpha_i \in \mathbb{R}^{d_i}$, $d$ being the dimension of the latent space. The modulator \acrshort{MLP} computes this parameter as follows:

\begin{equation*}
    \alpha_{i+1} = h^\prime_i = ReLU(w^\prime_{i+1}[h^\prime_iz]^T + b_{i+1})
\end{equation*}

From this formula, it can be noticed that the latent code $z$ is fed to each layer in a skip connection form. Another way to implement the modulation would be to concatenate the latent features with the input coordinates (like PointNet) and use it directly in the \acrshort{SIREN} network. However, this would only act as a phase-shift in the first layer, not having the sought effect on frequency and amplitude in all the layers.

The latent code itself can be produced by an autoencoder (third network) or an auto-decoder. In the first option, which is preferred, a continuous representation is built from the discrete signal. The second one, on the other hand, consists of randomly initialising the features and jointly optimize them with the network parameters.

In practice, a previous transformation is done. The signal is regularly partitioned, and each tile has its assigned latent code. This improves generalization by encoding these local, simpler parts, that have less variation than the global signal. To ensure the continuity of an image, for example, the tiles are also overlapped.

  \chapter{Adjoint Method}
\label{appenF}

Article \cite{m_bradley_pde-constrained_2019} presents a clear and general mathematical explanation of the adjoint method, which is going to be important to understand more specific cases.

The method arises to calculate the gradient in a \acrshort{PDE}-constrained problem. For example, given the
variables $x \in \mathbb{R}^{n_x}$ and the parameters $p \in \mathbb{R}^{n_p}$, there is a function $f(x, p): \mathbb{R}^{n_x} \times  \mathbb{R}^{n_p} \rightarrow \mathbb{R}$ it is wanted to know the gradient with regard to the parameters, i.e., $d_p f(x, p)$. This function is constrained, however, by a \acrshort{PDE} equation of the form $g(x, p) = 0$, where $g : \mathbb{R}^{n_x} \times \mathbb{R}^{n_p} \rightarrow \mathbb{R}^{n_x}$ and its derivative $g_x$ is everywhere non-singular.

This \acrshort{PDE} equation is normally calculated with the help of a software solver and may be discretized, as is the case of the CFD simulations - this is called the \textit{forward problem}. In this example, the program finds the values of the variables $x$ (field values) based on the values of the parameters of the simulations, such as boundary conditions or geometry. 

The function $f(x,p)$, in its turn, can be considered as a cost function and depends a lot on the goal of the engineer analyzing the problem. In general lines, it should use the value of $x$ found during the simulation and compute, based also on the parameters of the problem, as a measure of merit. Calculating the gradient of $f$ provides the possibility of minimizing it - the \textit{inverse problem}, used in Deep Learning during back-propagation.

The gradient $d_p f(x, p)$ can be easily approximated by finite differences. However, as this gradient requires the partial derivatives of the function w.r.t each parameter $p^i$, multiple simulations would have to be launched by changing the values of each $p^i$. Therefore, $n_p$ simulations would have to be launched, presenting itself as a very computationally expensive approach, especially in the case we propose in this document, using full mesh values as input.  

The adjoint method, however, presents an approach that allows the calculation of this gradient independent of the number of parameters. Solving the adjoint equation has usually the cost of solving only one simulation, leaving the whole procedure with a total cost of two simulations. This is, thus, ideal for our case.

\section{Mathematical definition}

For now, let's consider that the function $f$ depends only on the solution $x$ and not on the parameters $p$: $f(x)$. Two derivations can be done to get to the adjoint equation: direct, or using a Lagrangian approach. It is going to be described only the latter since it was preferred in the next article we are going to address. A \textit{Lagrangian} is normally used in optimization procedures to calculate the gradient when a constraint is imposed. It can be defined as the following:

\begin{equation*}
    \mathcal{L}(x,p,\lambda) = f(x) + \lambda^T g(x,p)
\end{equation*}
 Where $\lambda$ is the vector of \textit{lagrangian multipliers}. As the problem states that $g=0$, the value of $\lambda$ can be chosen freely, leading to:
 
\begin{figure}[ht!]
\includegraphics[width=10cm]{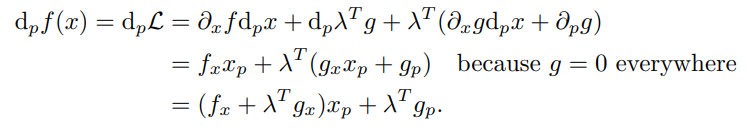}
\centering
\caption{Mathematical development of the adjoint}
\cite{m_bradley_pde-constrained_2019}
\end{figure}
 
Choosing then $g^T_x\lambda = -f^T_x$, the first term is zero. In fact, this is the adjoint equation, where $g^T_x$ is the matrix adjoint and $\lambda$ can also be called vector of adjoint variables. Finishing the calculation, it is possible to find the gradient:
 
 \begin{equation*}
     d_pf = \lambda^T g_p
 \end{equation*}
 
If $f$ also depends directly on the parameters $p$, which is the case in the next section, the formula gains another term:

\begin{equation*}
     d_pf = \lambda^T g_p + f_p
 \end{equation*}

As we can see from the formula above, the gradient is independent of the variables $x$, meaning it can be computed for an arbitrary number of parameters $p$ without the need to resolve additional simulations. How to calculate the terms of the equation above is a problem that will be addressed in the next sections, and requires, for the case of interest (\acrshort{CFD} simulations), the use of the \acrlong{NS} equations.

Lastly, it is important to notice the possibility of time-dependent problems, in which the function to be treated has the shape $F(x,p) = \int^T_0 f(x,p,t) dt$. It is constrained not only by the same condition as before but also from an \textit{ODE} in implicit form originating from a \acrshort{PDE} discretized in space, but not in time. The system becomes, thus, more complex to resolve.

\section{Adjoint Method in fluid dynamics}

In fluid dynamics, the adjoint method is commonly used in shape optimization. This information, also called sensitivity, can be calculated in a simulation and then used as a direction of descent for a boundary shape update in an optimization algorithm.

We will focus more on the part of calculating the gradient instead of the optimization, once the goal is to use it during training for the Deep Learning problem. In practice, this procedure can be done by a \acrshort{CFD} program, like OpenFoam.

The objective function can, for example in an inverse problem, be defined as the difference between the pressure calculated in the simulation ($p$) and a desired, \textit{ideal}, pressure on the surface of the geometry ($p_d$): $I = \frac{1}{2} \int_\mathcal{B} (p-p_d)^2 d \mathcal{B}$. This depends on the goal of the engineer. Many times, this is going to be a direct measurement of performance like the drag or lift coefficient.

This cost function $f(x,p)$, here takes the form of $I(w,\mathcal{F})$, where $w$ represents the flow-field variables (\ref{EulerEquations}) and $\mathcal{F}$ is the physical location of the boundary, therefore representing the shape. The design variables can, thus, be the geometrical position (x and y per point) or the initial flow conditions, such as the velocity in x and y. 

The constraint \acrshort{PDE} is represented by $R(w,\mathcal{F})=0$. Defining this time the Lagrangian multiplier as $\varphi$, the adjoint equation yields:

\begin{equation} \label{eq:adj1}
    \frac{\delta R}{\delta w}^T = \varphi \frac{\partial I}{\partial w}^T
\end{equation}
\begin{equation} \label{eq:adj2}
    \frac{\delta I}{\delta \mathcal{F}} = \frac{\partial I}{\partial \mathcal{F}} - \varphi^T \frac{\partial R}{\partial \mathcal{F}}
\end{equation}

\section{Discretization process of the equations}

The further development of the adjoint equations depends on the type of discretization chosen. Two options, very well explained in \cite{kenway_effective_2019}, arise:
\begin{itemize}
    \item Continuous: analytically derive the flow equations and then linearize the solution to solve them.
    \item Discrete: linearize the flow solution to solve it, and then derive the discretized equations.
\end{itemize}

\begin{figure}[htb]
\includegraphics[width=11cm]{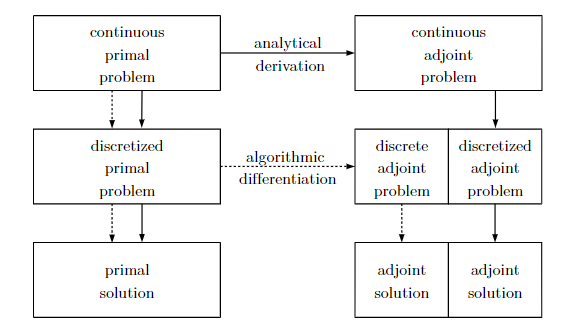}
\centering
\caption[Adjoint discretization scheme]
{Adjoint discretization scheme - illustrates the difference between the continuous and discretized adjoint approaches
 \cite{towara_discrete_2013}.}
\end{figure}

Both options have advantages and disadvantages. The first one, as an analytical method, represents the actual gradient equations and is a fast and low-memory implementation (implicitly forms the Jacobian), but requires hand derivation, being complex to develop and prone to human error; besides often including 
turbulence simplifications (frozen turbulence \cite{peter_numerical_2010}). Moreover, this approach is inaccurate for coarser meshes, being exact only at the limit at infinitely refined mesh. Lastly, as it uses the primal fields as a reference, coupling both systems (primal and adjoint), the primal solution has to be fully converged.

The discrete option, on the other hand, gives the exact gradient in each point/cell, independent of the coarseness of the mesh or the convergence and boundary conditions of the primal solution. It is also simpler, 
as it can use well-established numerical methods for the computation. However, this approach needs the explicit computation and storage of the Jacobian matix, in an way that is slower and demands a lot of memory. In addition, a non-converged primal solution will also not give the desired adjoint fields \cite{towara_discrete_2013}.

In any case, the discrete adjoint equations, as what happens with the flow, are linear systems. This linear system can be resolved with the traditional methods for solving this kind of problem, notoriously iterative mutigrid solution methods. Direct methods are too costly for real-world applications, where the inversion of the matrix for a 3d Cartesian and equally spaced grid ($n^3 elements$) is of complexity $n^7$.

\section{Deriving from discrete flow equations}

Article \cite{kenway_effective_2019} surveys different ways to apply this method in steady-state flow, also introducing a low-cost implementation. The resolution steps are (considering equations \ref{eq:adj1} \ref{eq:adj2}):

\begin{enumerate}
    \item Compute $\frac{\delta R}{\delta w}^T$ and $\frac{\partial I}{\partial w}^T$.
    \item Solve equation \ref{eq:adj1} to obtain $\varphi$.
    \item Compute $\frac{\partial I}{\partial \mathcal{F}}$ and $\frac{\partial R}{\partial \mathcal{F}}$.
    \item Solve equation \ref{eq:adj2}, finding thus, the sensitivity $\frac{\delta I}{\delta \mathcal{F}}$.
\end{enumerate}

The discrete adjoint computes those partial derivatives explicitly by numerical methods. Many methods are available, such as Finite difference (equation \ref{eq:finiteD}), Complex step - similar to Finite difference but uses complex perturbations, reducing truncation error -, and
Symbolic differentiation. Despite having good precision most of the time, these approaches are computationally inefficient for use in \acrshort{CFD} codes.

In contrast, \acrlong{AD} represents an efficient and accurate approach. The idea behind this procedure is that every computer code, no matter how complex it is, is reduced to elementary arithmetic operations by the machine. Therefore, the derivatives of each simple operation can be stored for a sequence and applied in the chain rule.

There are two ways to accumulate the derivatives: forward and reverse mode. Both will yield the same results; the only difference is the direction of accumulation, which can make one option more or less efficient than the other, depending on the case.

\begin{figure}[htb]
\includegraphics[width=11cm]{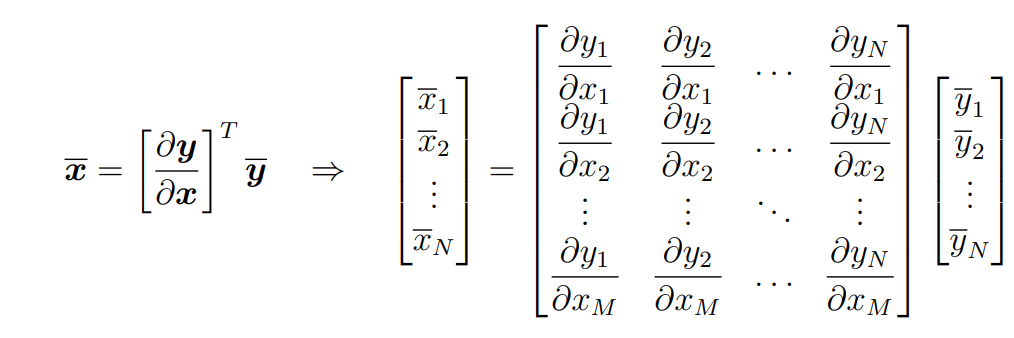}
\centering
\caption{Reverse mode AD}
Where $\boldsymbol{\Bar{y}} = [\Bar{y_1}, \Bar{y_2}, \cdots, \Bar{y_n}, \cdots, \Bar{y_N}] = [0, 0, \cdots, 1, \cdots, 0]$ \quad \quad
\cite{kenway_effective_2019}.
\label{ReverseAD}
\end{figure}

Forward \acrshort{AD} defines initially the input variables, and, as the code executes, it computes each derivative. Reverse mode does the contrary. The output variables are previously defined and the code executes, storing all intermediate values. The derivatives are, thus, calculated separately for each output variable. Figure \ref{ReverseAD}, shows how to compute the derivative vector of the output variable $y_n$: $\boldsymbol{\Bar{x}} = [
    \frac{\partial y_n}{\partial x_1}, 
    \frac{\partial y_n}{\partial x_2},
    \cdots,
    \frac{\partial y_n}{\partial x_M}
                        ]^T
                        $, where \textit{M} is the
number of input variables.

The derivatives in forward mode, however, are computed separately for each input variable $x_m$. Therefore, if there are more input than output variables ($M>N$), reverse mode should be used because it will do less matrix \textit{x} vector product operations, and vice-versa. An efficient code should leverage this property to apply each of the options when necessary.

As the discrete approach is computationally and memory expensive, due to the above operations, it is important to well optimize the AD. Another way to do this is to explore the sparsity of the partial derivatives matrices, not being necessary to calculate all of their components. 

In practice, there are two ways of implementing \acrshort{AD} (independent of the mode) in a code. The first is by transforming the source code by adding extra statements, which will compute derivatives on the fly and can be almost as efficient as the continuous approach. However, it cannot be applied in object-oriented languages, like C++. That's why implementations that use Openfoam for the primal solver, like DAFoam \cite{he_dafoam_2019} and \cite{towara_discrete_2013}, use \textit{operator overloading}. This method creates a new data type to overload the original operations and functions, rendering the code 2 to 4 times slower and at least twice the memory consumption, as it stores both values and derivatives.

DAFoam is an Openfoam based framework for shape optimization. It uses the discrete adjoint for \textit{Free Form Deformation} (FFD) \cite{sederberg_free-form_1986} - a volumetric approach where control points are defined to be displaced considering the gradient of the objective function and its constraints, assuring a feasible geometry \cite{anderson_parametric_2012}. It is suited for small and medium deformations and parametrizes the geometry change instead of the geometry itself \cite{kenway_cad-free_2010}.

\section{Deriving from the original flow equations}

A lot of the advancement of this approach was done Jameson from the late 80s to the start of the 21st century. In this section, details are given from his article \cite{jameson_aerodynamic_2003}, using control theory. More recent works include \cite{kavvadias_continuous_2014, kavvadias_proper_2015, zymaris_continuous_2009, papoutsis-kiachagias_continuous_2014}, which are the base for the Openfoam's implementation.

The development of the terms of equation \ref{eq:adj2} depends on \textit{R} - the \acrshort{PDE}s being used in the discretization of the simulation. A simpler case of the Navier-Stokes equations, which do not include the viscous terms are the Euler equations:

\begin{figure}[ht!]
\includegraphics[width=6cm]{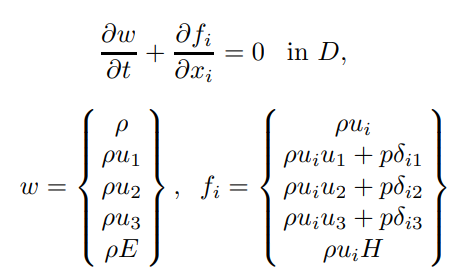}
\centering
\caption[Euler equations]
{Euler equations - Cartesian coordinates and velocity components are represented by x1, x2, x3 and u1, u2, u3; summation over i = 1 to 3 is implied by a repeated index i. In the order, there are the continuity equation, 3 momentum equations and the energy equation \cite{jameson_aerodynamic_2003}.}
\label{EulerEquations}
\end{figure}

\subsubsection{Representation of the Physical Boundary}
The parameters $p$ of the solution above were introduced as the physical representation of the boundary, $\mathcal{F}$. This concept can be a little hard to grasp. In practice, the solution is mapped to a fixed computational domain with coordinates $\xi_1, \xi_2, \xi_3$ 

This change of coordinates allows to pass the parameters from  $\mathcal{F}$ to $S$, and changes the whole resolution of the problem. In the finite volume discretized representation, $S$ is a matrix whose elements are the face areas of the computational cells projected in the x1, x2, and x3 directions. It is $S$, then, that is used in the functions $I$ and $R$. Other modifications are also done to ensure that the gradient that does not depend on the mesh deformation itself, but on the true boundary movement and flow solution.

Lastly, it is important to notice that the presented approach focused on laminar flow. When turbulent equations enter, the resolution becomes even more complex and further mathematical development can be done \cite{kavvadias_continuous_2014}.
  \chapter{Fluid equations}
\label{appenG}

The high Reynolds number used in the simulations implicates a turbulent regime, which is characterized by an unsteady and chaotic flow with the presence of eddies. The numerical solution can be achieved by directly discretizing the Navier Stokes equations \ref{NvSkSimplified}, in what is called \gls{DNS}. However, to capture the small nuances of the turbulence, a very fine mesh is necessary, resulting, in most cases, in a prohibitively expensive simulation.

Different strategies arose to deal with this issue; one of them being the \gls{LES}. This method reduces the computational cost by filtering the small-scale fluctuations of the flow, focusing only on the larger vortexes. Nevertheless, this comes at the cost of worse resolution, especially near the wall, where the boundary layer causes small-scale phenomena.

Another alternative, that is largely used, is the \gls{RANS} equations, which modify the original equations by averaging the velocity and pressure fields over time. 

\begin{equation*}
    u = \Bar{u} + u' \quad, p = \Bar{p} + p'
\end{equation*}

In this way, the resolution of the system becomes simplified and quicker. To close the set of \acrshort{RANS} equations for the turbulent regime, it is necessary to find the turbulent viscosity $\nu_t$. Different assumptions can be done for this goal. A simple, and yet very common model, is the Spalart-Allmaras \cite{kostic_review_2015}, which has only one equation that considers a linear behavior of the referenced viscosity \cite{nasa_langley_research_cente_turbulence_nodate}. This approach is less modern and can sometimes yield important errors. It is, however, well studied and was the first model to have an Openfoam adjoint implementation.

A more modern model is the $K-\omega \: SST$ \cite{menter_ten_2003}, which combines the advantages of the $k-\omega$ near the wall, with the $k-\epsilon$, switching between both. The three of them also consider linear viscosity but use 2 equations. Its Openfoam adjoint implementation became available in version 2206.

\end{document}